\documentclass[twocolumn]{aastex62}

\usepackage{amsmath,natbib, float, color, soul, xspace}
\usepackage[normalem]{ulem}
\newcommand*\Msolarh[0]{h^{-1} \, \mathrm{M_{\odot}}}
\newcommand*\msolarh[0]{h^{-1} \, \mathrm{M_{\odot}}}
\newcommand*\Mpch[0]{h^{-1}\,\mathrm{Mpc}}
\newcommand*\Gpch[0]{h^{-1}\,\mathrm{Gpc}}

\newcommand*\kms[0]{\mathrm{km \, s^{-1}}}

\newcommand*\vcut[0]{$v_{\mathrm{cut}}$}

\newcommand*\model[0]{$\sigma_8$-$\Omega_m$}
\newcommand*\mathmodel[0]{\sigma_8, \, \Omega_m}

\newcommand*\sig{\textsc{$\sigma_8$}\xspace}
\newcommand*\Sig{\textsc{$\mathcal{S}_8$}\xspace}
\newcommand*\om{\textsc{$\Omega_m$}\xspace}

\newcommand*\hecs{\textsc{HeCS-SZ}\xspace}
\newcommand*\sqdeg{\mathrm{deg}^2\xspace}

\newcommand*\ysz{$Y_\mathrm{SZ}$\xspace}
\newcommand*\planck{\textit{Planck}\xspace}

\newcommand{\possessivecite}[1]{\citeauthor{#1}'s \citeyear{#1}}

\newcommand*\new[1]{{#1}}

\newcommand*\param[1]{#1}
\newcommand*\result[1]{#1}

\begin{document}

\correspondingauthor{Michelle Ntampaka}
\email{michelle.ntampaka@cfa.harvard.edu}

\author{Michelle Ntampaka}
\affiliation{Center for Astrophysics $|$ Harvard \& Smithsonian, Cambridge, MA 02138, USA}
 \affiliation{Harvard Data Science Initiative, Harvard University, Cambridge, MA 02138, USA}

\author{Ken Rines}
\affiliation{Department of Physics \& Astronomy, Western Washington University,  Bellingham, WA 98225, USA}

\author{Hy Trac}
\affiliation{McWilliams Center for Cosmology,  Department of Physics, Carnegie Mellon University,  Pittsburgh, PA 15213, USA}

\title{Cluster Cosmology with the Velocity Distribution Function of the HeCS-SZ Sample }

\begin{abstract}
We apply the Velocity Distribution Function (VDF) to a sample of Sunyaev-Zel'dovich (SZ)-selected clusters, and we report preliminary cosmological constraints in the $\sigma_8$-$\Omega_m$ cosmological parameter space.  
The VDF is a forward-modeled test statistic that can be used to constrain cosmological models directly from galaxy cluster dynamical observations.  The method was introduced in \cite{2017ApJ...835..106N} and employs line-of-sight velocity measurements to directly constrain cosmological parameters; it is less sensitive to measurement error than a standard halo mass function approach.  
 The  method is applied to the Hectospec Survey of Sunyaev-Zeldovich-Selected Clusters (\textsc{HeCS-SZ}) sample, which is a spectroscopic follow up of a \planck-selected sample of 83 galaxy clusters.  
 Credible regions are calculated by comparing the VDF of the observed cluster sample to that of mock observations, yielding \result{\smash{$\Sig \equiv \sigma_8 \left(\Omega_m/0.3\right)^{0.25} = 0.751\pm0.037$}}.  These constraints are in tension with the \planck Cosmic Microwave Background (CMB) TT fiducial value, which lies outside of our 95\% credible region, but are in agreement with some recent analyses of large scale structure that observe fewer massive clusters than are predicted by the \planck fiducial cosmological parameters.
\end{abstract}

\keywords{cosmology: cosmological parameters --- galaxies: clusters: general --- methods: statistical}

\section{Introduction}
\label{sec:intro}

Galaxy clusters are massive, gravitationally bound collections of hundreds to thousands of galaxies.  The standard cosmological model predicts that the abundance of clusters as a function of their mass and redshift depends sensitively on the underlying cosmological parameters. Because cluster counts depend on the cosmological model, cluster abundance measurements can be used to put constraints on cosmological parameters such as the amplitude of matter fluctuations, \sig, and the matter density parameter, \om \citep[e.g.][]{1998ApJ...504....1B}.  More recently, forward-modeling approaches have been used to describe cluster abundance, not by the cluster counts as a function of mass and redshift, but by the distribution of direct observables \citep[e.g.][]{2012PhRvD..86l2005W, 2014arXiv1411.8004H, 2016MNRAS.462.4117C, 2017A&A...607A.123P, 2017ApJ...835..106N}.   Forward-modeling approaches can minimize biases that would otherwise be introduced by mass measurement error and can provide a complementary and mass-independent way to evaluate the abundance of clusters.

Cluster abundance and other large scale structure (LSS) measurements can be used to constrain cosmological models, but some current LSS data are in tension with \planck CMB constraints of cosmological parameters.  
Notably, SZ surveys find fewer massive clusters than are predicted by the \planck CMB fiducial cosmology \citep[e.g.][]{2016A&A...594A..24P}; one interpretation of this is that the 
\planck SZ survey prefers a \sig lower than the CMB fiducial cosmology. While some LSS probes are consistent with \planck constraints \citep[e.g.][]{2015MNRAS.446.2205M},  others prefer a low \sig{} \citep[e.g.][]{2013MNRAS.432.2433H, 2017MNRAS.465.1454H}.  In a rigorous analysis of WiggleZ, SDSS RSD, CFHTLenS, CMB lensing and SZ cluster count by  \cite{2017arXiv170809813L}, these LSS probes are found to be inconsistent with \planck CMB cosmological constraints. 
This tension may be due to unaccounted systematic errors or it may be indicative of more interesting physics that needs to be included in the model.

A number of explanations have been suggested to resolve this tension.
X-ray cluster mass measurements under the assumption of hydrostatic equilibrium can incorporate nonthermal pressure support in the form of a bias parameter, $b$, that relates the SZ mass to the true cluster mass, $M_\mathrm{SZ}=(1-b)M_\mathrm{true}$.  SZ cluster masses calibrated on X-ray observations must also take this bias factor into account.  When this bias factor is allowed to be large, the tension between CMB and SZ parameters is significantly reduced; the mass bias required to bring cluster observations into agreement with the CMB anisotropy is $b\approx0.42$ \citep{2016A&A...594A..24P}.
Adding a non-zero neutrino mass can also reduce the tension, but at the same time increases tension in other parameters predicted by \planck CMB and SZ.  Notably, it lowers the \planck constraints for the Hubble constant  \citep{2016A&A...594A..24P}.  Though a number of explanations have been suggested to resolve this tension, the source of the disagreement is not yet agreed-upon within the community.

{Dynamical cluster mass estimates provide a complementary way to explore the LSS-CMB tension}.  Dynamical mass estimates utilize the virial theorem, relating the LOS velocity dispersion of cluster members, $\sigma_v$, to cluster mass, $M$, as a power law.  This approach famously led to \possessivecite{zwicky1933rotverschiebung} inference of dark matter in the Coma cluster.  Hydrodynamical simulations find that the velocity dispersion of dark matter particles are similar to those of galaxies  \citep[e.g.][]{2010ApJ...708.1419L}, though selecting only the brightest galaxies in a cluster tends to bias the galaxy velocity dispersion low \citep{2013ApJ...772...47S, 2013MNRAS.436..460W}.  Because the dynamical mass method provides a relatively unbiased probe of cluster masses that is complementary to other cluster mass estimates, virial-theorem-based approaches are used in modern cluster mass estimates \citep[e.g.][]{2010ApJ...721...90B, 2010ApJ...715L.180R, 2013ApJ...772...25S, 2014ApJ...792...45R, 2015ApJ...799..214B}.  {Cluster counts as a function of velocity dispersion have also been proposed as a method for constraining cosmological parameters \citep{2016MNRAS.462.4117C}.}

This work utilizes the Hectospec Survey of Sunyaev-Zeldovich-Selected Clusters \citep[\hecs,][]{2016ApJ...819...63R}.  The \hecs sample is an MMT/Hectospec spectroscopic follow up of a sample of 83 clusters selected from \planck observations.  {In their analysis of the \hecs sample, \cite{2016ApJ...819...63R} find that these clusters are dynamically colder than expected, having smaller velocity dispersions, $\sigma_v$, than is predicted by a \planck-selected sample of clusters for the \planck fiducial cosmology.}   A mass bias together with a velocity bias of galaxies could explain the discrepancy.  However, the biases required to remove the tension between CMB and cluster cosmological parameters would need to be large, in some cases, outside of the range presented in recent literature.  \cite{2016ApJ...819...63R}  conclude that another explanation may be necessary to resolve the tension.  

{To explore the LSS-CMB tension, we utilize cluster line-of-sight velocity observations of the \hecs sample in conjunction with a forward modeling dynamical approach.}  The cluster sample is analyzed using the Velocity Distribution Function (VDF), which was first introduced in \cite{2017ApJ...835..106N}.  This method utilizes a forward-modeled test statistic that quantifies the abundance of clusters by measurements of their dynamics, through comparing summed {probability density functions (PDFs) of clusters' measured line-of-sight (LOS) velocities to that of a mock catalog.  }

We present a summary of the cluster observations,  the mock observations, and the VDF methodology in Section \ref{sec:methods}, results and constraints on \sig and \om in Section \ref{sec:results}, and a discussion of the results in Section \ref{sec:discussion}.

\section{Methods}
\label{sec:methods}

\begin{figure*}
	\begin{center}
	\begin{tabular}{c c}
		\includegraphics[width=0.5\textwidth]{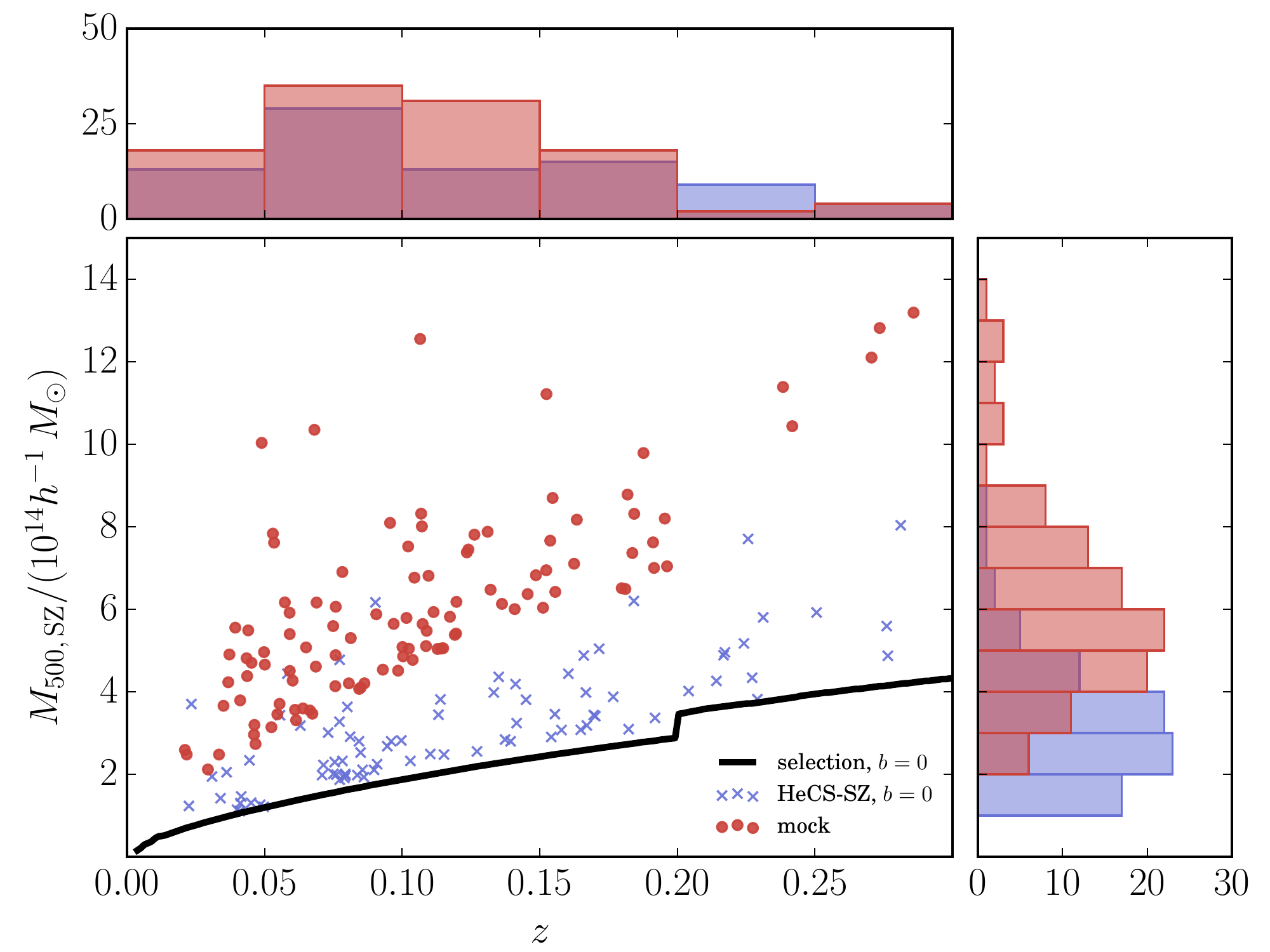} & \includegraphics[width=0.5\textwidth]{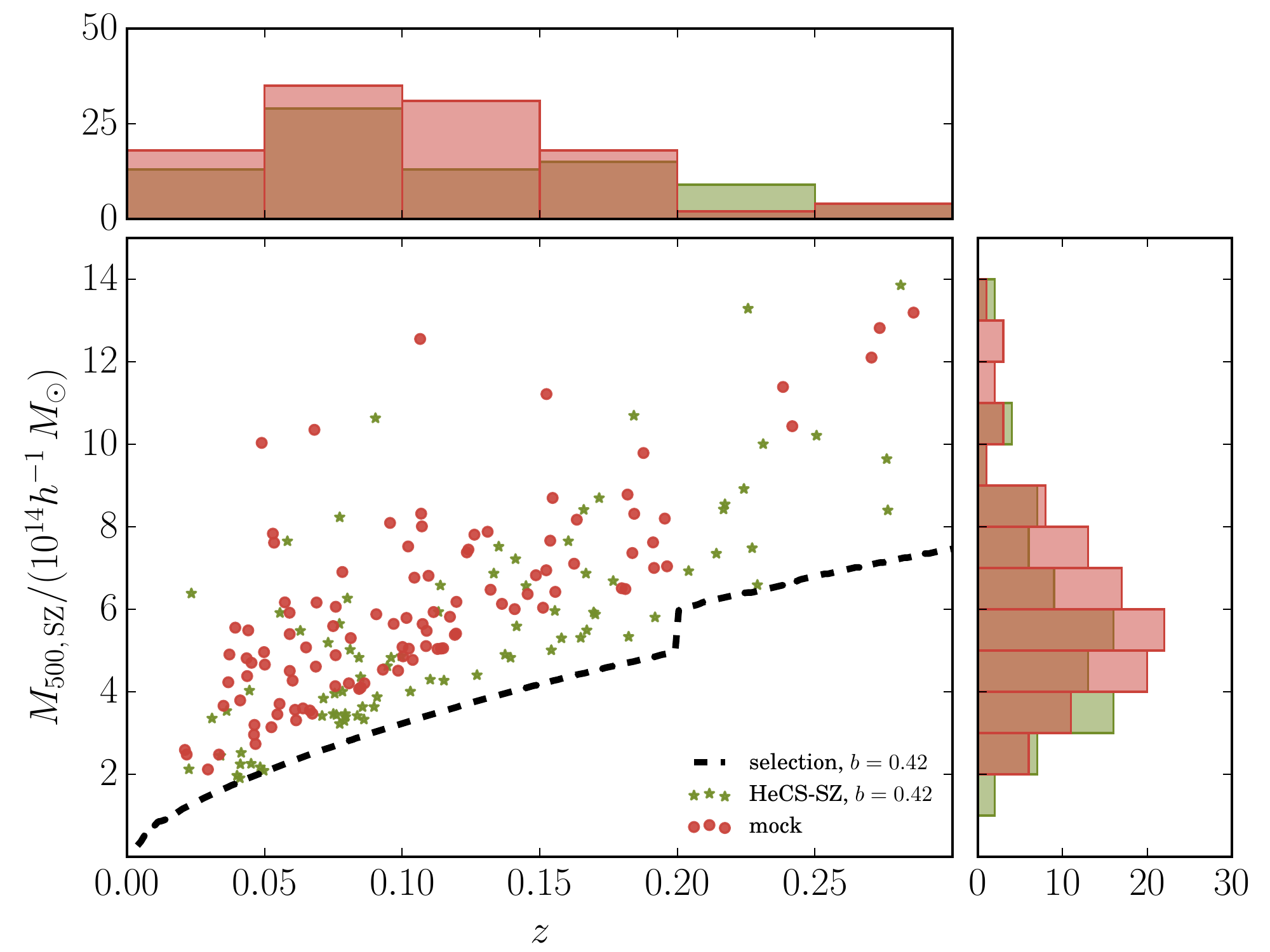} \\
		\end{tabular}
	\caption{{Left:  The redshift and SZ mass ($M_\mathrm{500,\,SZ}$) distribution of one sample mock observation (red circles) and that of the \hecs sample with $b=0$ (blue x's). The selection function with $b=0$ (solid black curve) shows the limit above which {the sample is modeled to have integral completeness $\mathcal{C}\in[0.6, 1.0]$, with $\mathcal{C}=0.8$ shown}.  The discontinuity at $z=0.2$ is due to two different observing methods being employed above and below this redshift.  When the selection function is applied to the mock light cones, it is varied by a multiplicative factor {of $(1-b)^{-1}$} until the number of clusters above the selection function matches the number of clusters expected, given the \hecs observation and choice of $\mathcal{C}$.  The mock observation is clearly not in agreement with that of the \hecs observation for $b=0$.  
	Right:  \cite{2016A&A...594A..24P} reports a bias of $b=0.42$ is needed to bring the \planck observed $M_{500, \mathrm{SZ}}$ masses into agreement with the mass distributions predicted by the \planck CMB fiducial cosmology.   When this  bias is applied to the  \hecs observation (green stars) and selection function (black dash), the \hecs sample is in better agreement with the mock observations, which assume the \planck fiducial cosmology.  Summary PDFs of mass and redshift representative of the entire suite of mock observations are shown in Figure \ref{fig:masshist}.  }}
       	\label{fig:catcomp}
      	\end{center}
\end{figure*}

\subsection{HeCS-SZ}
\label{sec:hecs}

The Hectospec Survey of Sunyaev-Zeldovich-Selected Clusters \citep[\hecs,][]{2016ApJ...819...63R} is an SZ-selected sample of clusters  \citep{2014A&A...571A..29P}.  {The full \planck-selected sample contains 87 clusters selected above the 80\% completeness limit of the \planck medium-deep sky or the 50\% completeness limit of the \planck shallow sky \citep{2014A&A...571A..29P};} \hecs comprises 83 clusters selected randomly from these 87 clusters.  The sample ranges from redshift \param{0.05} up to \param{0.3} and utilizes SDSS DR6 Legacy and SDSS DR10 imaging as well as MMT/Hectospec spectroscopic follow up of potential cluster members. {When two high-error galaxies are excluded, the remaining 25,112 galaxies in the \hecs sample have mean line of sight velocity error of \param{$\approx 30\,\kms$}.}

The {cluster} selection function of the {\hecs} sample is shown in Figure \ref{fig:catcomp}.  As redshift increases, so does the minimum mass a cluster must have to be detected by \planck; this is due to the \planck beam size.  At high $z$ and low $M$, the SZ signal is diluted by the large beam size, causing these clusters to have low signal to noise.  \param{The selection function is discontinuous at $z=0.2$ because the \hecs sample utilizes observations from two SDSS data releases. The sample for $z<0.2$ utilizes observations from SDSS DR10 and has an effective sky area of $11589 \, \sqdeg$ and a selection function that corresponds to \planck's 80\% completeness limit for the medium-deep sky.  The sample for $0.2 < z < 0.3$ utilizes imaging observations from SDSS DR6 Legacy and spectroscopic observations from the Hectospec Cluster Survey \citep{2013ApJ...767...15R}.  It has an effective sky area of $8417 \, \sqdeg$ and the completeness limit jumps by 20\% compared to the $z<0.2$ sample.}

Figure \ref{fig:catcomp} shows the selection function of the \hecs observation, as well as the redshift and SZ-determined $M_{500}$ values of the 83 clusters in the \hecs sample.  {The Planck medium-deep survey zone comprises 41.3\% of the full sky, while the shallow survey zone comprises 56\% (including the region around the Galactic Plane).  Along the selection function shown in Figure \ref{fig:catcomp}, the likelihood of a cluster being detected by \planck is 80\% for the medium-deep survey zone and 50\% for the shallow survey zone.  This likelihood increases with increased cluster SZ mass, $M_{500,\,\mathrm{SZ}}$, therefore clusters lying above the selection function are more likely to be detected (See \cite{2014A&A...571A..29P} for complete details of the cluster selection function, survey zones, and completeness).  To account for the varying completeness in this range, we define an integral completeness,  
\begin{equation}
\mathcal{C}=\frac{N_\mathrm{observed}}{N_\mathrm{true}},
\label{eq:completeness}
\end{equation} 
the ratio of the number of \planck-detected clusters above the selection function to the true number of clusters in the sky and redshift regions of interest.  We adopt a conservative flat prior of $\mathcal{C} \in [0.6, 1.0]$ on the integral completeness of the cluster sample.}

The \hecs $M_{500}$ shown in this figure are calculated from the $Y_\mathrm{SZ}$ signal and reported by \cite{2014A&A...571A..29P}.  Figure \ref{fig:catcomp} shows results for two sample values of the mass bias $b$, which parameterizes the scaling between a cluster's true mass and the observed $M_\mathrm{SZ}$ mass as $M_\mathrm{SZ} = (1-b) M_\mathrm{true}$.   A bias parameter of $b=0.42$ is needed to bring SZ cluster masses into agreement with \planck CMB  \citep{2014A&A...571A..29P, 2016ApJ...819...63R}, and employing this bias effectively increases the $M_{500}$ values of the \hecs clusters reported on the vertical axis of Figure \ref{fig:catcomp}.  
 See \cite{2016ApJ...819...63R} for further details of the \hecs observations.

A simple cylindrical cut is used to select apparent cluster members, including both true cluster members and interloping field galaxies, from the full \hecs observation.  This cylindrical cut uses angular extent and velocity cuts to correspond with the size and velocity dispersion of a cluster with \param{$M_\mathrm{200c} \geq 1\times10^{15} \msolarh$}  at the \planck fiducial cosmology:  a comoving \param{$1.6\Mpch$ outer aperture, initial velocity cut of $2500 \,\kms$ about the cluster center and no $\sigma_v$ velocity paring}, according to the method detailed in \cite{2017ApJ...835..106N}.  

The location of the cluster center in the plane of the sky and the location of the cluster velocity center are both chosen iteratively.  A cylinder with \param{$R_\mathrm{aperture} = 1.6\Mpch$  and  $|v|\leq2500 \,\kms$}  is centered initially on the SZ center and redshift of the cluster reported by \cite{2014A&A...571A..29P}.  {The plane of sky center of mass and the mean velocity are calculated and the cylinder is recentered on this new location.  The cylinder center iteratively moves in this manner until convergence (defined as \param{$\Delta R < 0.02 \Mpch$ and $\Delta v < 50 \,\kms$}); most clusters converge in one or two iterations.}

The central region of the \hecs observation is then removed; the resulting cylindrical selection has a \param{$0.25\Mpch$ axial hole} through the center.  {The velocity bias between substructure and dark matter particles is small in the outer regions of clusters but increases toward the center, making the outer region a better probe of the cluster's dynamical state (Aung et al., \textit{in prep.}).}  Furthermore, there may be systematic differences between the mock catalogs and \hecs clusters due to the high concentration of galaxies in this region stemming from fiber collisions, observational selection effects, or distance resolution in the $N$-body simulation used to build the mock observation that destroys substructure near the center of simulated cluster.  

Determining physical comoving distances to apply the fixed aperture requires some knowledge of the underlying cosmological parameters, and the measurement is particularly sensitive to \om.  We note that varying \om by $\approx 10\%$ changes the effective aperture by $\lesssim 2\%$.  Because this effect is relatively small, we opt to adopt the straightforward approach of assuming the \planck fiducial cosmology and a \param{$1.6 \Mpch$} comoving aperture for the \hecs clusters.

\subsection{Mock Observations}

\begin{figure*}[t]
	\begin{center}
	\begin{tabular}{c c}
		\includegraphics[width=0.5\textwidth]{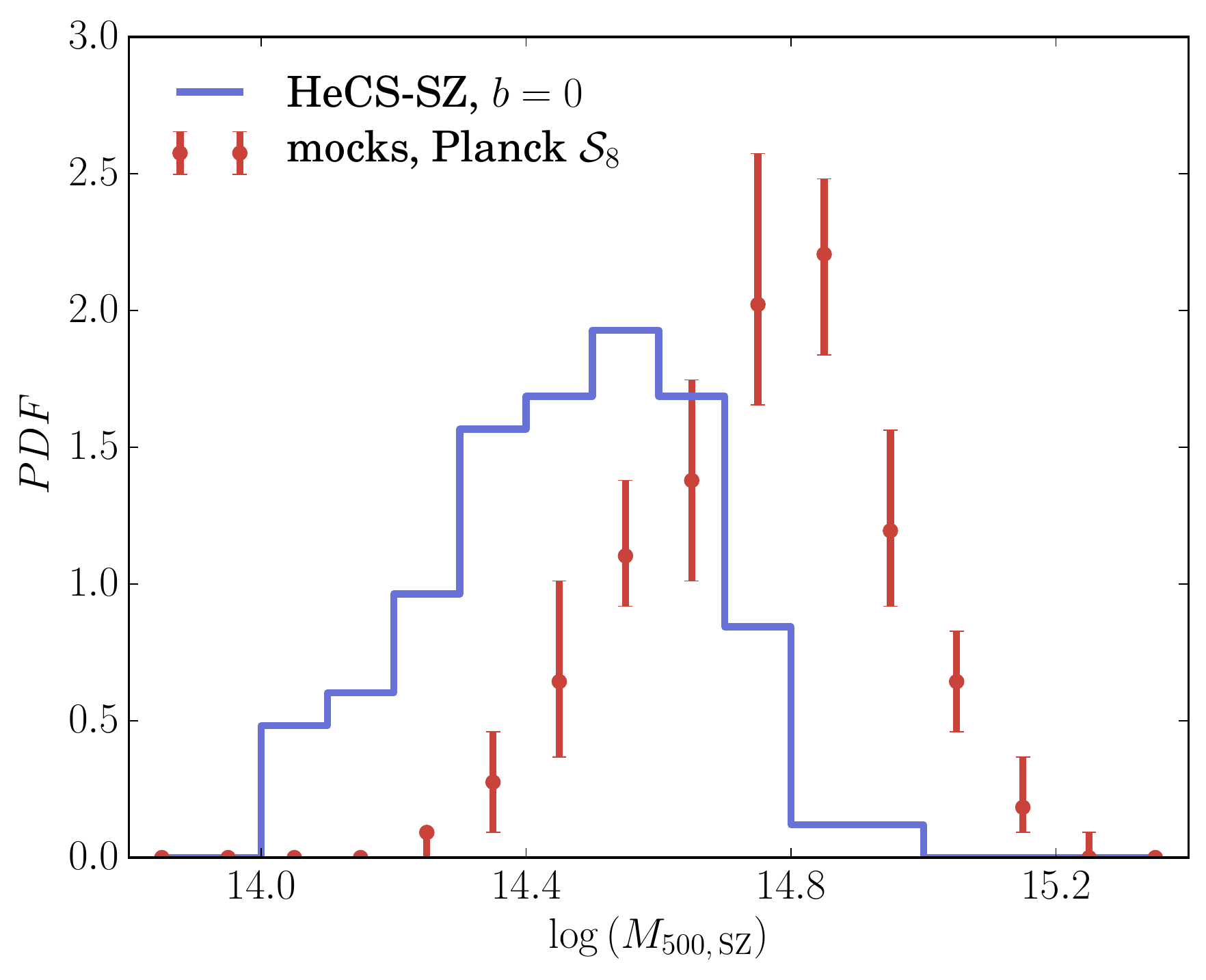} &  \includegraphics[width=0.5\textwidth]{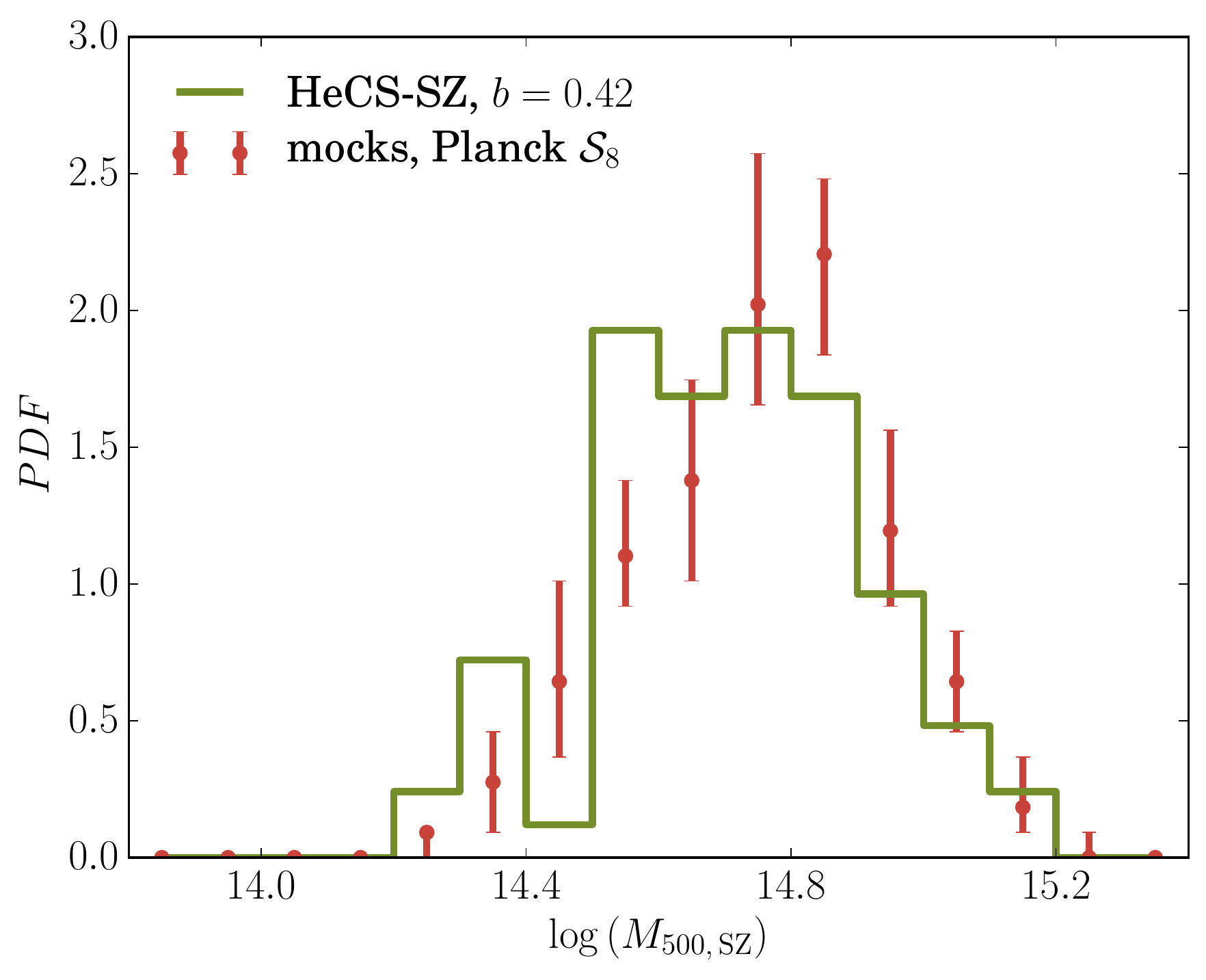} \\
		\includegraphics[width=0.5\textwidth]{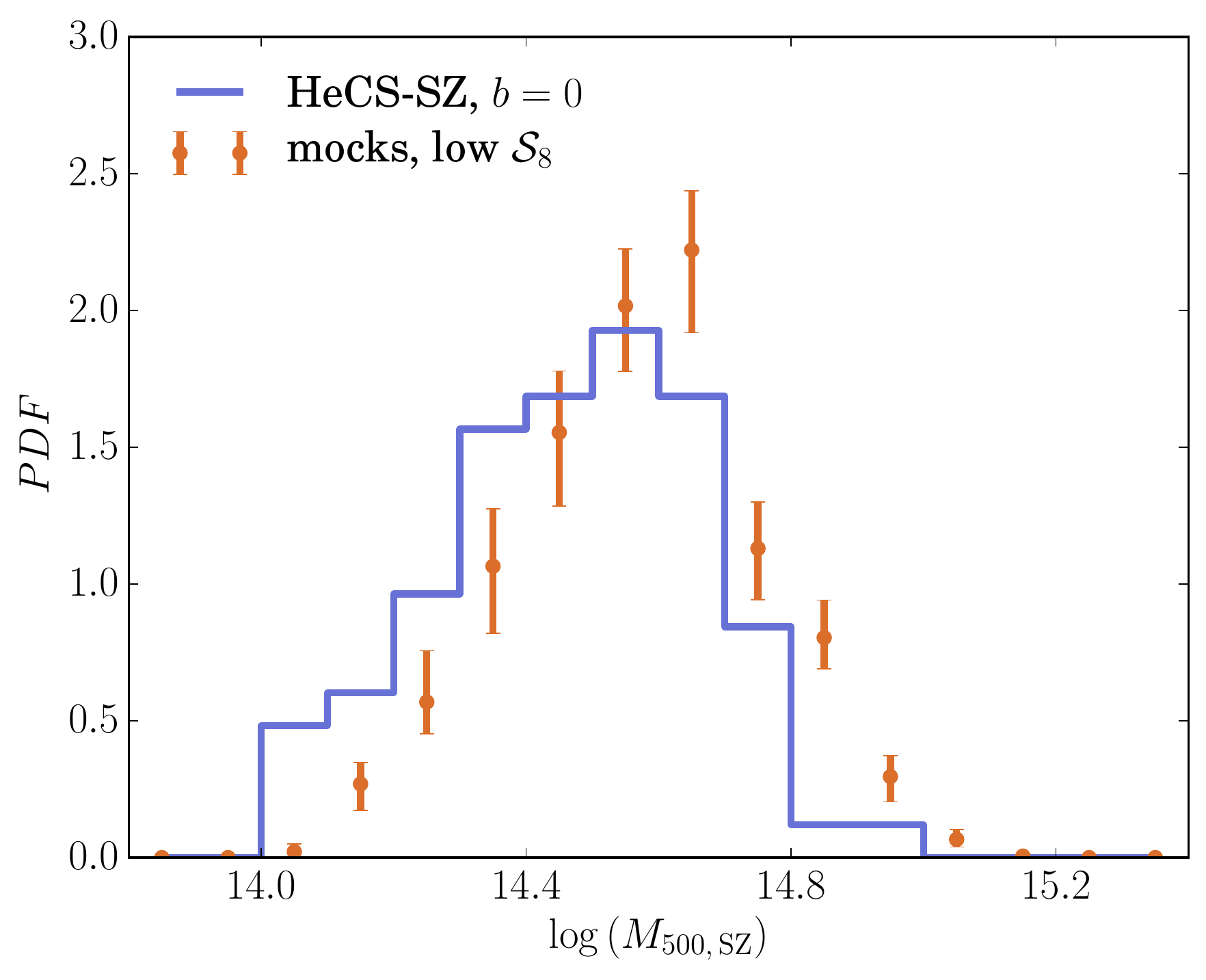} & \includegraphics[width=0.5\textwidth]{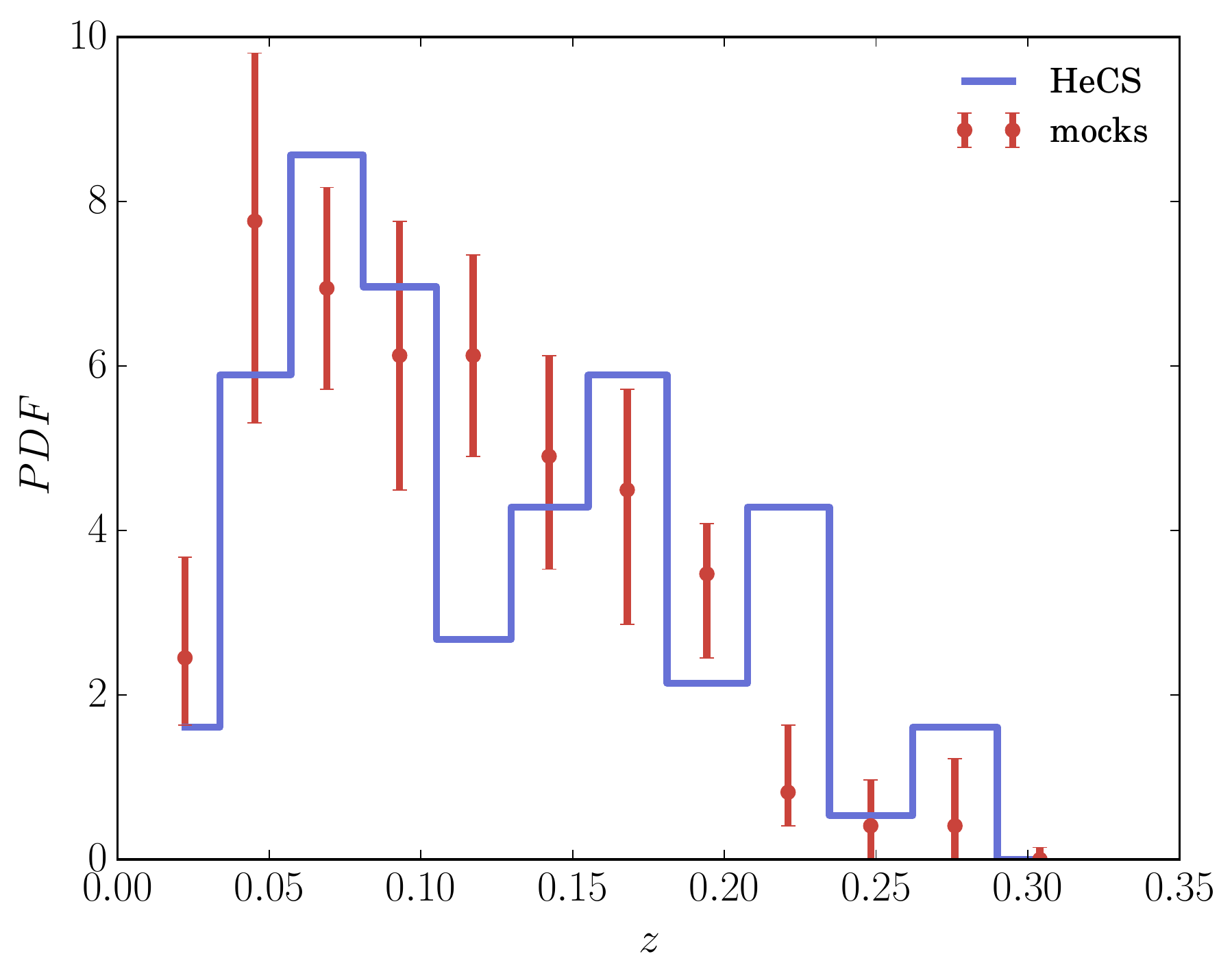} \\
	\end{tabular}
	
	\caption{Top left:  A PDF of cluster SZ masses, $M_\mathrm{500,\,SZ}$ in units of $\msolarh$,  assuming no mass bias.  The \hecs  masses (blue solid) are as reported by \planck.  Mock observation masses (red error bars showing the middle 68\% of the mock observation PDF) include scatter in the $Y_\mathrm{SZ}(M_\mathrm{500,\,SZ})$ relationship to properly model Eddington bias.  
	Top right:  A significant bias must be applied to the \hecs cluster masses ($b=0.42$, green solid) to bring these into agreement with the masses predicted by the mock observations (red error bars) because the mock observations are built on a simulation that assumes the Planck CMB fiducial cosmology. 
	Bottom left:  Alternately, invoking a low \Sig can bring the unbiased \hecs observations (blue solid) into agreement with mock observations (orange error bars); $\Sig=0.70$ is shown for illustrative purposes.
	Bottom right:  The range of redshift distributions in the fiducial cosmology mock observations (red error bars) and the \hecs catalog (blue solid) have similar redshift distributions.  {For brevity, $\mathcal{C}=0.8$ is shown.}}
       	\label{fig:masshist}
      	\end{center}
\end{figure*}

The mock observations are created from UniverseMachine \citep{2018arXiv180607893B} galaxies incorporated into the publicly available Multidark MDPL2 $N$-body simulation\footnote{http://www.cosmosim.org/} \citep{2016MNRAS.457.4340K}.  The MDPL2 simulation uses a $\Lambda$CDM cosmology, with cosmological parameters consistent with the \planck CMB fiducial cosmology \citep{2014A&A...571A..16P}:  $\Omega_{\Lambda} = 0.693$, $\Omega_m = 0.307$, $\Omega_b = 0.048$, $h = 0.678$, $n_s=0.96$, and $\sigma_8 = 0.823$.  

The MDPL2 simulation contains $3840^3$ particles in a box with comoving length $1.0h^{-1}\rm{Gpc}$ with mass resolution of  $1.51\times10^9\msolarh$. 
Halo and subhalo properties are reported from the publicly available Rockstar halo catalog \citep{2012ascl.soft10008B}, and halo masses  reported here are with respect to the critical density $\rho_\mathrm{crit}$.

UniverseMachine is an empirical model of the merger history and star formation history of galaxies.   The model paints galaxies onto halos and subhalos in the Multidark simulation using halo merger trees;  halo mass, halo assembly history, and redshift all inform galaxy star formation rate. UniverseMachine tracks orphan galaxies, which are galaxies for which the hosting substructure has been numerically disrupted in the simulation, allowing for richer cluster mock observations than would be possible with the $N$-body simulation alone.  All UniverseMachine galaxies with stellar mass $\geq 10^{9.5}\msolarh$ are used to construct the mock observations.  {Because the \hecs sample has an average measurement error of \param{$\approx 30\,\kms$}, each galaxy in the mock catalog is given a velocity error selected from a Gaussian with width \param{$30\,\kms$}}.  For further details about the UniverseMachine galaxy catalog, see  \cite{2018arXiv180607893B} and references therein.

The UniverseMachine galaxies in comoving $1.0h^{-1}\rm{Gpc}$ snapshots for twelve redshifts ranging from $z=0.0$ to $z=0.304$ are used to construct \param{eight} full-sky light cones.  The light cones differ only in the location at which the observer is placed.
Each full-sky light cone is used to construct \param{ten} mock observations.  These mock observations are modeled to follow the \hecs observation closely:  they adopt the redshift range \param{($0.05<z<0.3$)} and sky area ($11589 \, \sqdeg$ for $z<0.2$ and $8417 \, \sqdeg$ for $0.2 < z < 0.3$) of \hecs.  

\ysz signals are assigned to each cluster with \param{$M_{200} \geq 10^{14.2} \msolarh$}  according to the power law relation from \cite{2016A&A...594A..24P}, which assumes the universal pressure profile of \cite{2010A&A...517A..92A}.  The power law relation is employed with bias parameter \param{$b=0$} and lognormal scatter $\sigma_{\ln{Y}}$ included.  The $Y_\mathrm{SZ}(M_\mathrm{SZ})$ relationship is inverted, and $M_\mathrm{SZ}$ values assigned to each cluster based on the cluster's $Y_\mathrm{SZ}$ signal.  Because this process includes scatter $\sigma_{\ln{Y}}$, it properly forward-models the sample of clusters which would be realistically observed and, therefore, accounts for Eddington bias.

\begin{figure}[b!]
	\begin{center}
		\includegraphics[width=0.45\textwidth]{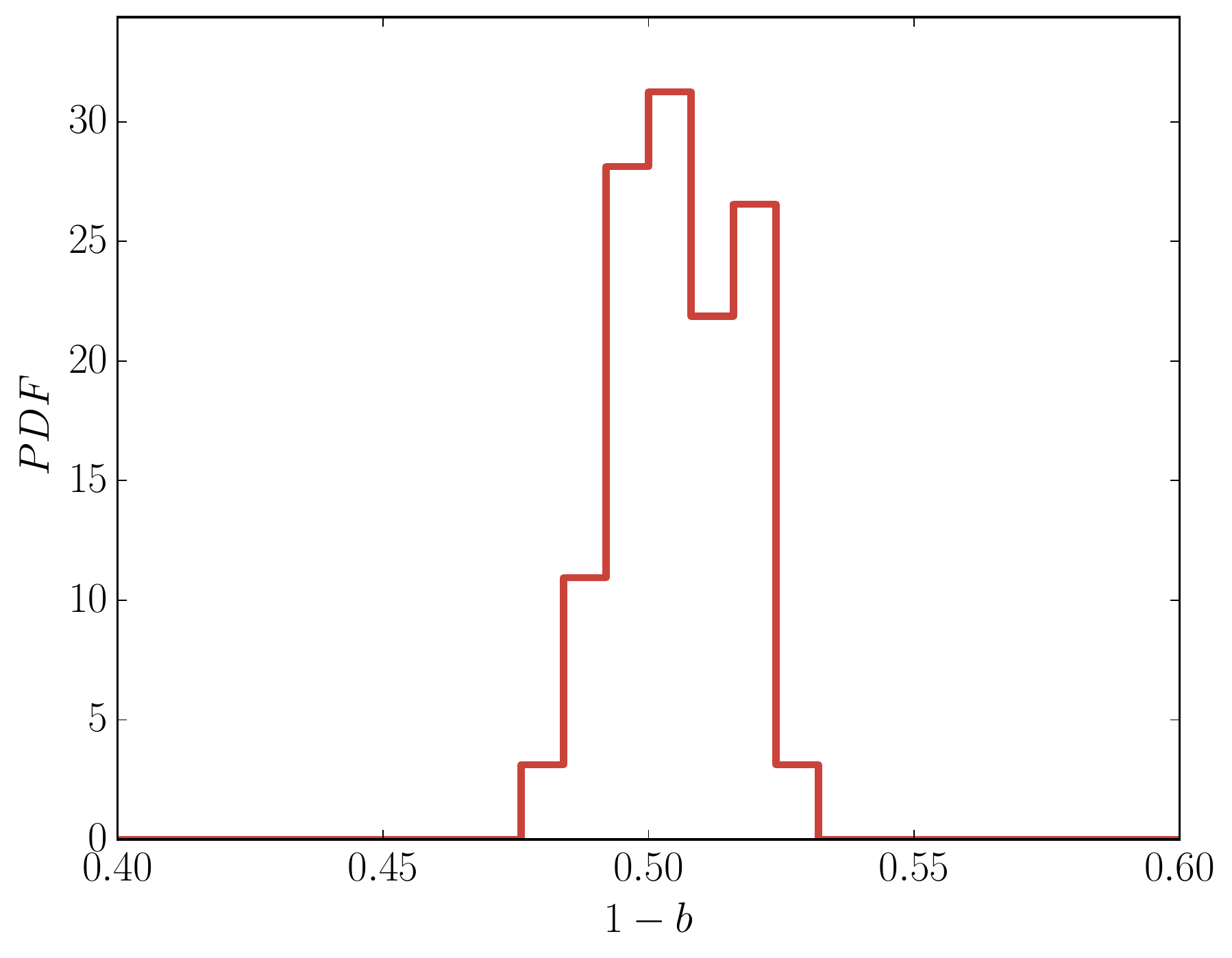} \\
	\caption[bias]{A PDF of bias, $b$, needed to scale the selection function (shown in Figure \ref{fig:catcomp}) to match the expected cluster counts for $\mathcal{C}= 0.8$.  A significant bias is necessary to bring the observed cluster masses into agreement with that of a simulation at the \planck CMB fiducial cosmology.  The VDF provides a way to explore this tension by utilizing cluster dynamical measurements. }
       	\label{fig:biashist}
      	\end{center}
\end{figure}

Apparent cluster members, including both true cluster members and interloping field galaxies, are selected with the same method described in Section \ref{sec:hecs}:  a simple cylindrical cut with a \param{$1.6\Mpch$} aperture, initial velocity cut of \param{$2500 \,\kms$} about the cluster center, and with \param{no $\sigma_v$ clipping}.  The cluster centers on the plane of the sky, as well as in velocity space, are determined by the same \param{iterative scheme} described in \ref{sec:hecs}, and the central \param{$0.25\Mpch$} is removed from the sample.  

Figure \ref{fig:masshist} shows a PDF of mock catalog and \hecs sample SZ masses and redshift.  For $b=0$, the distribution of masses disagrees significantly, but invoking a mass bias of $b=0.42$ can bring the mass distributions into agreement.  Alternately, a mock catalog with a low value of \Sig can also bring the mass distributions of the observed and simulated clusters into agreement.  The redshift distribution of of the \hecs sample agrees with that of the mock catalog.  

{Mock observations for both the fiducial cosmology as well as for the nonfiducial cosmologies} are calculated as in \cite{2017ApJ...835..106N}.  Briefly, each cluster is given a weight according to the cluster's redshift, $z$, and  mass, $M_{200}${, both of which are known quantities that are extracted from the simulation}.  This weight, $w$ is given by
\begin{equation}
w=\frac{\mathrm{HMF}(M, z, \sig{}, \om)}{\mathrm{HMF}(M, z, \sig{}_\mathrm{, sim}, \om_\mathrm{, sim})}
\end{equation}
{the ratio of the nonfiducial analytic halo mass function (HMF) to the simulation HMF for the full $1.0 \Gpch$ box at the most appropriate redshift snapshot.  For the analytic HMF we use that of \cite{2008ApJ...688..709T}} with an average halo density of \param{$200\rho_\mathrm{crit}$.  Further details about the HMF of the Multidark suite of simulations can be found in \cite{2017MNRAS.469.4157C}.}

{The selection function shown in Figure \ref{fig:catcomp} is scaled by a multiplicative factor  until the clusters above the selection have $\sum w = 87 \mathcal{C}^{-1}$, where 87 is the number of clusters in the \planck-selected sample from which the 83 \hecs clusters were randomly selected and $\mathcal{C}$ is the integral completeness above the selection function.  }

{The multiplicative scaling of the selection function of the mock observations relates to the bias factor that may be applied to the \hecs observations, $b$, {as $(1-b)^{-1}$}.  Therefore, at the fiducial cosmology, a multiplicative scaling of $1.72$ applied to the scaling relation shown in Figure \ref{fig:catcomp} is equivalent to the bias $b=0.42$} that is needed to bring \hecs cluster observations into agreement with the \planck CMB TT fiducial cosmology.  Figure \ref{fig:biashist} shows a histogram of biases needed to scale the simulation at the fiducial cosmology to match expected cluster counts for $\mathcal{C}=0.8$.  Biases that are slightly larger than the \planck-reported value are needed to bring the mock observations into full agreement with the observed cluster counts.  This is due to the fact that the simulation uses the cosmological model at the central point of the \planck fiducial cosmology; simulations at lower \Sig cosmologies, but still within the CMB TT constraint  contours, would prefer lower biases.

In the case where cluster mass is defined by an average halo density with respect to the critical density, e.g.~the $200\rho_\mathrm{crit}$ definition used in this work, it is unlikely that the $M(\sigma_v)$ relation changes dramatically with changing \om \citep[e.g.][]{Evrard:2008aa, 2016MNRAS.456.3068O}.  Therefore, for nonfiducial, non-simulated cosmologies, cluster velocity distributions for individual clusters \new{are} assumed to remain unchanged.

{
\subsection{Mock Observations with Systematics}
\label{sec:altcat}
In addition to the standard method outlined in Section \ref{sec:methods}, we explore how systematic differences between the mock catalog and the \hecs sample might affect the resulting  \Sig constraints.  We construct {eleven} mock catalogs with systematics to explore this.  {The first four catalogs explore changes in parameters applied to both the \hecs and mock catalog, while the remaining seven catalogs explore systematic changes to the mock catalogs only.}
\begin{enumerate}
	\item \textbf{Large Aperture}:  To explore how the choice of cylinder size may bias the results, a large cylinder with $R_\mathrm{ap}=2.3\Mpch$, $v_\mathrm{cut}=3785\kms$, and $R_\mathrm{hole}=0.25\Mpch$ is used to select galaxies in both the \hecs sample and also in the mock catalog.  The $R_\mathrm{ap}$ and $v_\mathrm{cut}$ values are selected to correspond to the typical radius and $2\sigma_v$ of a $3\times10^{15}\Msolarh$ cluster.
	\item \textbf{No Axial Hole}:  To explore how the choice of axial hole may bias the constraints, no axial hole is removed from the cylinder center.  All other parameters of the standard catalog remain unchanged.  This change is applied to both the \hecs sample and also to the mock catalog.
	\item \textbf{Sigma Clipping}:  To reduce the effects of interlopers, 2-sigma iterative velocity clipping is applied both to the \hecs sample and also to the mock catalogs.  Additional details about this iterative clipping can be found in \cite{2017ApJ...835..106N}.
	\item \textbf{Low Redshift}:  To explore whether the constraints are driven by the high-mass sample at high redshift, these clusters are removed from the sample.  Both the \hecs sample and also the mock catalogs are limited to samples with $z\leq0.2$.	
	\item \textbf{Y-M Scatter}:  To explore whether an underestimation of the scatter in the Y-M scaling relation may bias the results, the level of scatter in the Y-M scaling relation is increased by 50\%.  This change is applied only to the mock catalog to assess for a potential systematic difference between the mock and \hecs samples.  The increase in scatter effectively makes the average mass of the mock sample smaller by scattering low-mass clusters above the selection function shown in Figure \ref{fig:catcomp}.
	\item \textbf{Radial Selection}:  To explore whether small differences between the \hecs sample and the mock observation in the the radial distribution function (RDF, discussed in detail in Section \ref{sec:rdf}) bias the results, the mock clusters are subsampled to match the \hecs RDF.
	\item \textbf{Rich Clusters}:  To explore how the choice of richness may bias the constraints, only rich clusters, defined as having at least $40$ galaxies, are included in the mock catalog.
	\item \textbf{Reduced Interlopers}:  To explore whether an overabundance of interlopers may bias the results, the threshold mass for interlopers is increased to $M_{200}\geq10^{12}\Msolarh$.  This change is applied only to the mock catalog to assess for a potential systematic difference between the mock and \hecs samples.  The increase in interloper mass cut effectively decreases the number of interloping galaxies in the mock catalog.
	\item \textbf{Red Fraction}:  \hecs galaxies are preferentially selected to lie on the red sequence and therefore are more likely to be cluster members (and less likely to be interlopers) than a stellar-mass-selected sample.  To test for a bias from preselecting likely cluster members, we select galaxies from the mock catalog with a similar bias towards cluster members. This is done by subsampling true cluster members to 80\% of the original galaxy population and interloping field galaxies to 20\% of the original galaxy population.  This change is applied to the mock sample only.	
	\item {\textbf{Quiescent Galaxies}:  \hecs galaxies are preferentially selected to lie on the red sequence, while the sample in the standard catalog is mass-selected.  Differences are found between the velocity dispersions of red galaxies versus the velocity dispersions of all galaxies in a cluster sample, with velocity dispersions of a color-selected sample mildly biased compared to than that of a stellar-mass-selected sample \citep[e.g.][]{2002A&A...387....8B, 2005MNRAS.359.1415G, 2006A&A...456...23B, 2013ApJ...773..116G, 2017MNRAS.468.1824O, 2018MNRAS.478.2618F, 2018MNRAS.481.1507B}.   To test for a bias stemming from differences in galaxy selection, we select galaxies from the mock catalog with specific star formation rate (sSFR) $\leq\,10^{-2}\,\mathrm{Gyr}^{-1}$ \citep[as in, e.g.,][]{2015MNRAS.446..521S}.  This change is applied to the mock sample only.}
	\item {\textbf{Biased Velocities}:  To further explore how velocity bias may affect the resulting constraints, \new{we artificially impose a velocity bias} on the galaxies in the standard mock catalog, reducing the velocity of every galaxy in the mock catalog to $0.95$ times its true value.  It should be noted that this check likely overemphasizes the true velocity bias in two ways:  first, by imposing the bias on all galaxies in the sample, including interlopers, and second, by failing to disentangle the fact that this bias tends toward zero for well-sampled clusters \citep{2013ApJ...772...47S}.  This change is applied to the mock sample only.}\label{item:biasedv}

\end{enumerate}
Unless otherwise noted, all parameters are identical to the standard catalog and the method is applied to both the \hecs sample as well as to the mock catalog.  Table \ref{table:summary} summarizes the key details of these mock catalogs with systematics.  }

\subsection{The Velocity Distribution Function}

The velocity distribution function \citep[VDF,][]{2017ApJ...835..106N} is a forward-modeled test statistic that can be used to compare distributions of observed cluster member velocities to those predicted with simulations.   The VDF can be used directly to explore constraints on cosmological parameters and is less affected by measurement errors associated with dynamical masses or the resulting Eddington bias \citep{1913MNRAS..73..359E}.  The VDF, $dn(v)/dv$, is the sum of {probability distribution functions (PDFs) of galaxy LOS velocities} and is given by
\begin{equation}
	\frac{dn}{dv} (v) = \left[\frac{1}{N}\sum_{i=1}^{N} \left[ \mathrm{PDF}(|v|) \right]_i \right]_{A, z_\mathrm{min}, z_\mathrm{max}},
	\label{eq:VDF}
\end{equation}
where $\mathrm{PDF}(|v|)$ denotes a probability distribution function of the absolute value of galaxy LOS velocities, the index $i$ denotes a sum over $N$ clusters (83 for the \hecs sample and {$87 \times \mathcal{C}^{-1}$ for the mock observations}), $A$ indicates that the VDF is calculated for a given sky area, and $z_\mathrm{min}, z_\mathrm{max}$ indicates that the VDF is calculated for a given redshift range\footnote{Note that the VDF takes a slightly different form than the one presented in  \cite{2017ApJ...835..106N}.  Here, we have removed the volume element, which is sensitive to the underlying cosmological parameters, and instead defined the test statistic by limits on parameters that are invariant under changes in cosmology: sky area, redshift range, and the number of clusters observed.}.  {We use {six} velocity bins of width \param{$\Delta v=250\,\kms$, {with edges}  from $0$ to $1500\,\kms$}.}

\begin{figure}[]
	\begin{center}
	\includegraphics[width=0.5\textwidth]{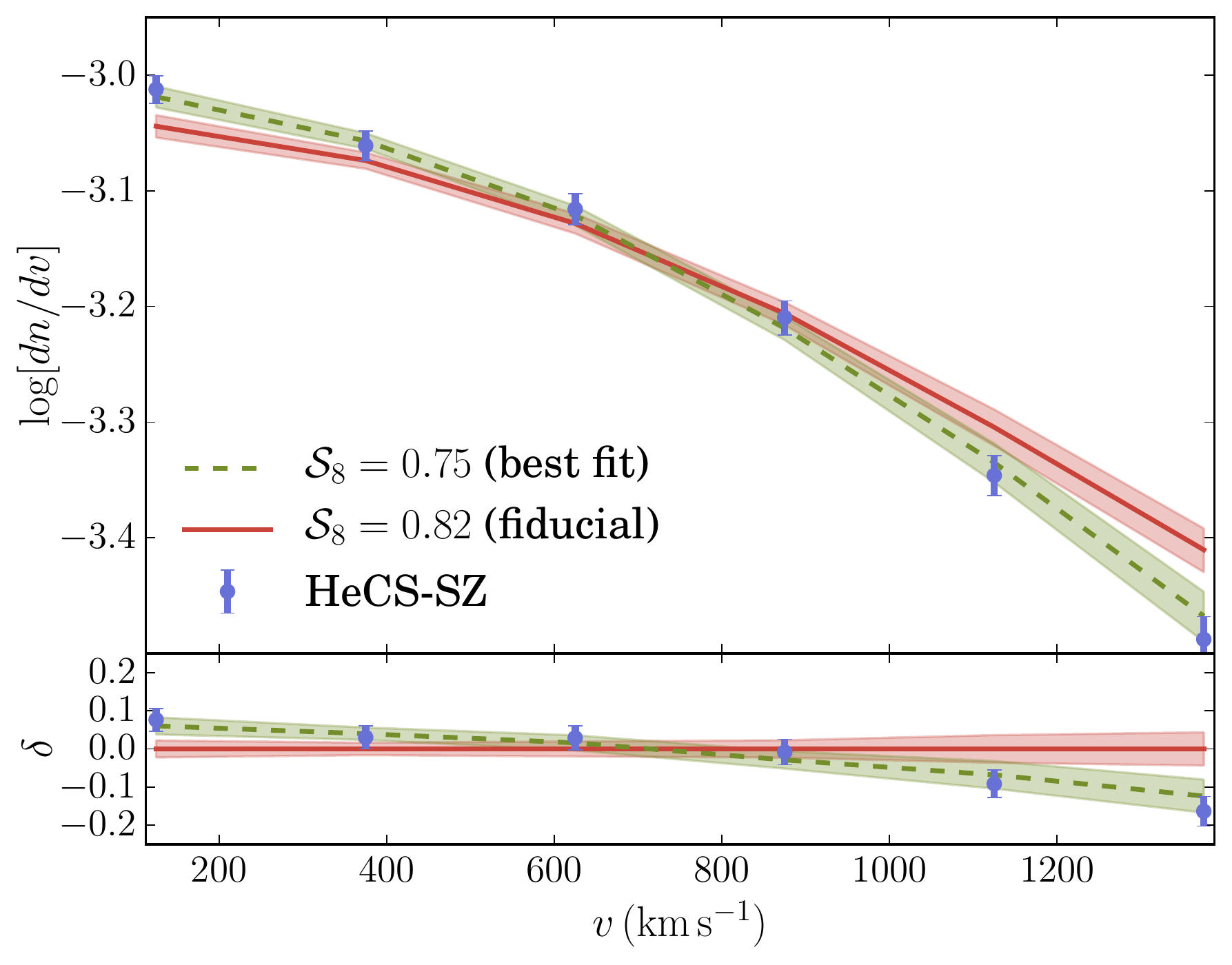}
	\end{center}
	\caption{{Top panel: The VDF of the mock observations at two different cosmologies (colored curves with band indicating the middle 68\% of mock observations), for {the standard catalog with} integral completeness $\mathcal{C}=0.8$.  There is a dearth of high-velocity members in the \hecs catalog (blue points with Poisson error bars)  compared to the fiducial cosmology ($\Sig=0.82$, red solid), but a lower \Sig  cosmology (e.g.~$\Sig=0.75$, green dash) is more consistent with the data.  This agrees with the results of the top left panel of Figure \ref{fig:masshist}, which shows that unless one invokes a significant mass bias, there are fewer high-mass clusters in the \hecs observation than in the mock observations.
	Bottom panel:  The fractional difference between models, $\delta$, is the largest at very high and very low velocities, where the technique has the most resolving power.  Near $\approx 800 \kms$, $\delta$ crosses 0, indicating that bins near these central velocities are not as useful in differentiating models.
	}
	}
	\label{fig:vdf}
\end{figure}

The resulting VDF is shown in Figure \ref{fig:vdf}.  As in \cite{2017ApJ...835..106N}, the VDF has the most resolving power at high and low velocities.  The crossover point where velocity bins have no resolving power has shifted downward from $\approx 1200 \, \kms$ in \cite{2017ApJ...835..106N} to \result{$\approx 800\, \kms$}; this is due to the changes in the cluster selection method and effective volume compared to the previous tests on mock catalogs.  

The VDF in Figure \ref{fig:vdf} shows that the \hecs clusters are dynamically colder than those in the simulation at the fiducial cosmology, since the VDF has a dearth of high-velocity members compared to the mock observation at the fiducial cosmology.  {As was noted in \cite{2016ApJ...819...63R}, the \hecs clusters have a smaller $\sigma_v$  than expected from a \planck-selected cluster sample at the \planck CMB fiducial cosmology.  This could be caused by a number of factors, including a true dearth of high-mass clusters or a bias between velocity dispersions in the simulations and those of observed clusters.}

\begin{figure*}[]
	\begin{center}
	\begin{tabular}{c c c}
	Standard & Large Aperture & No Axial Hole \\
	\includegraphics[width=0.3\textwidth]{fig4.pdf}&\includegraphics[width=0.3\textwidth]{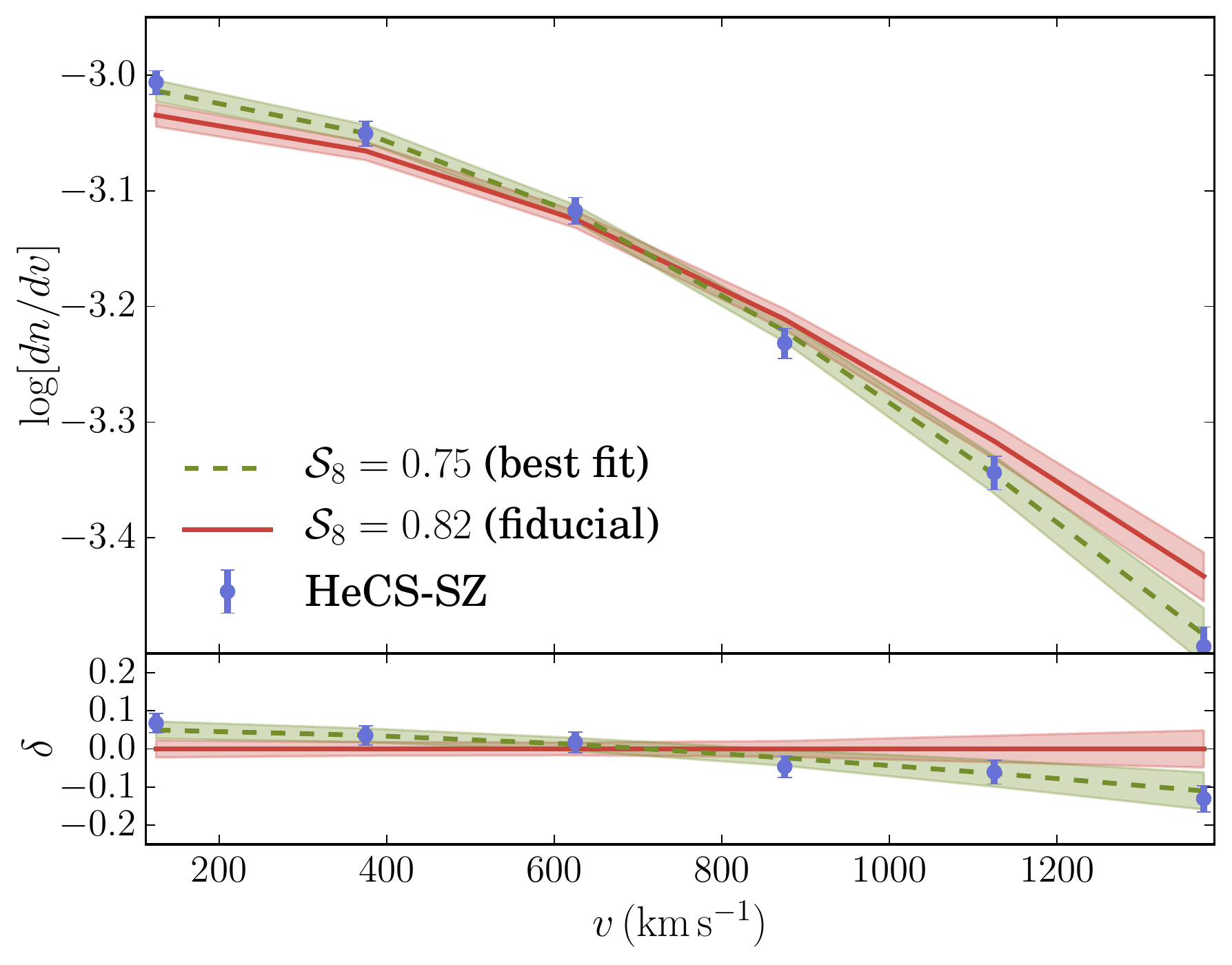}&\includegraphics[width=0.3\textwidth]{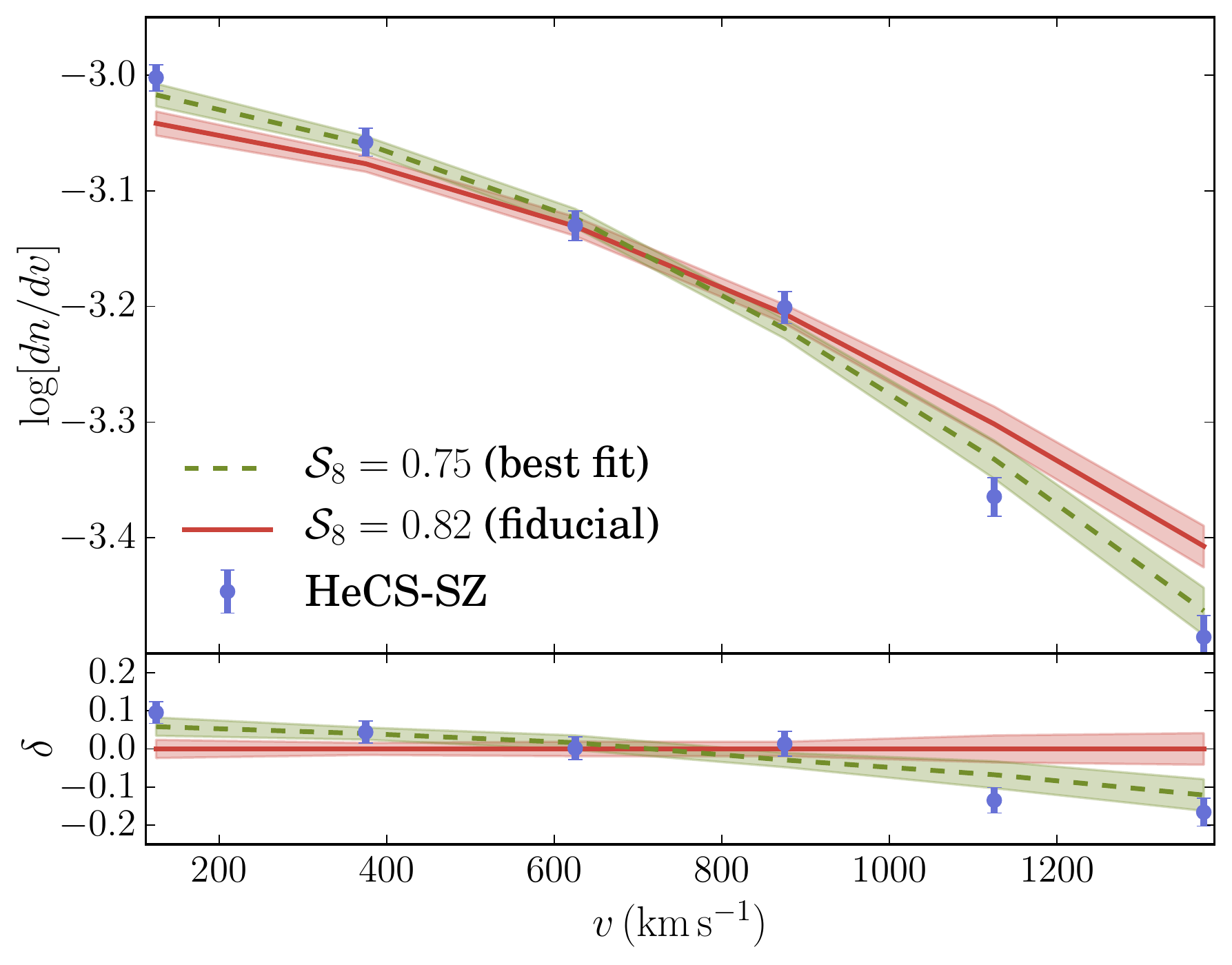}\\[5ex]
	Sigma Clipping & Low Redshift & Y-M Scatter\\
	\includegraphics[width=0.3\textwidth]{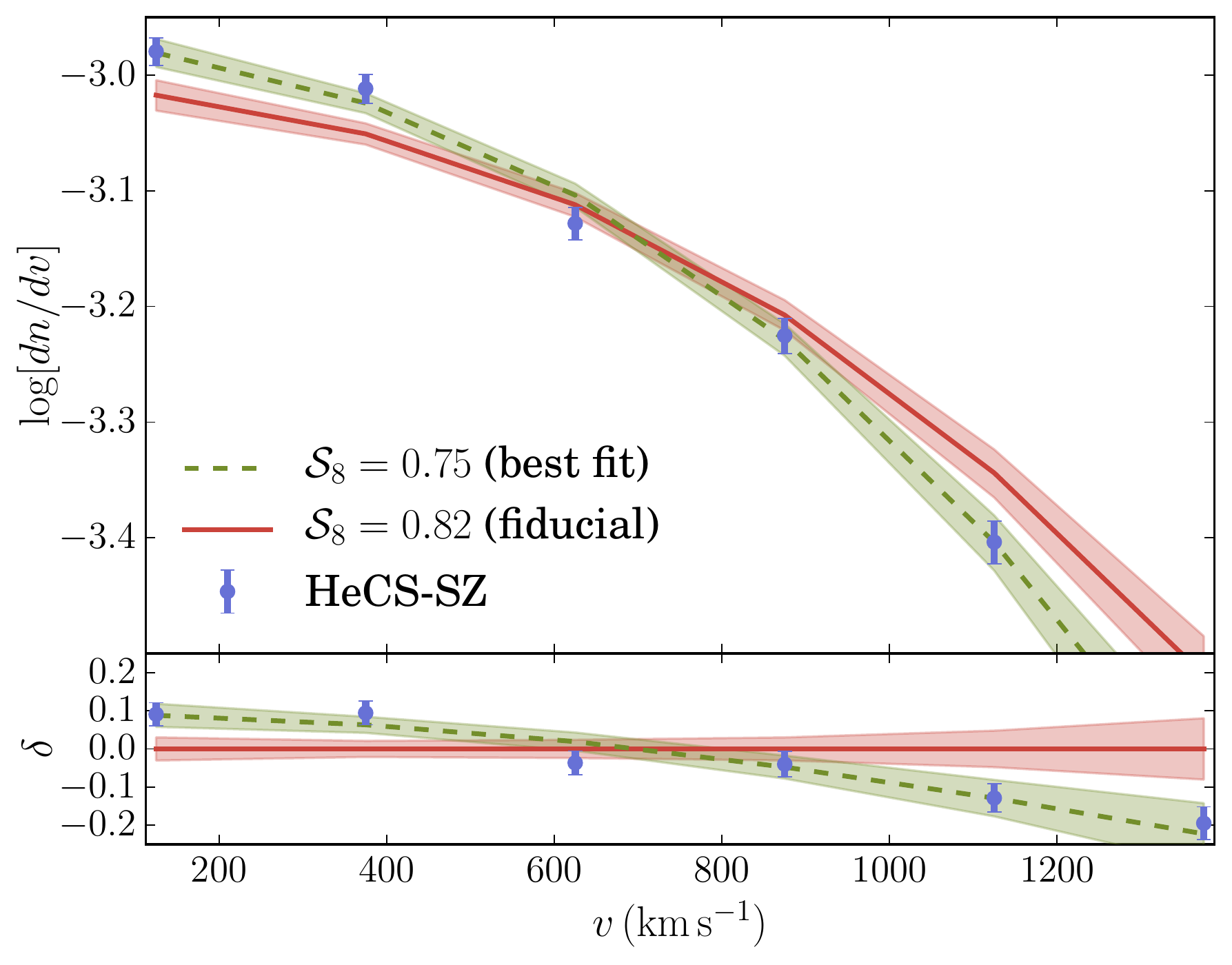}&\includegraphics[width=0.3\textwidth]{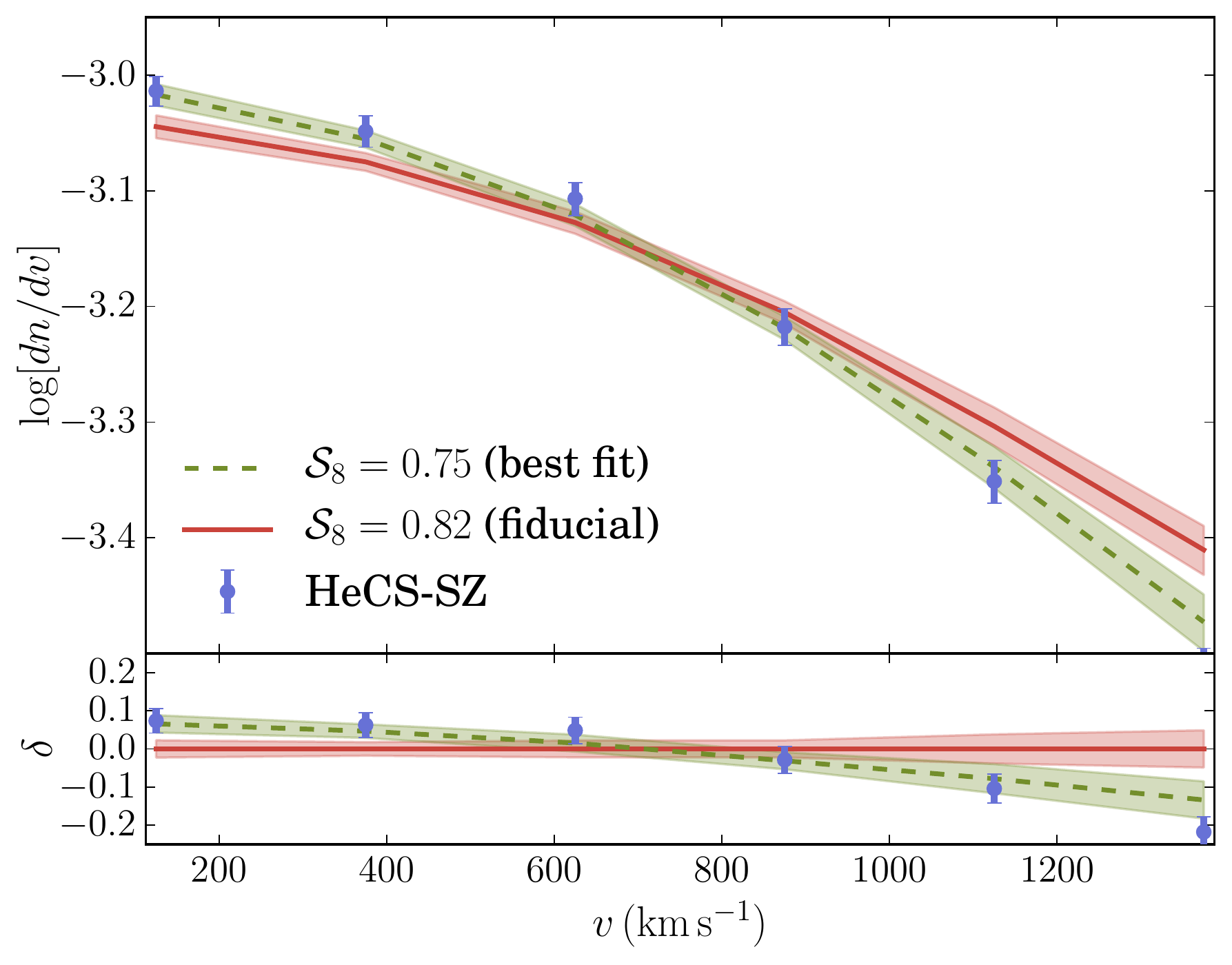}&\includegraphics[width=0.3\textwidth]{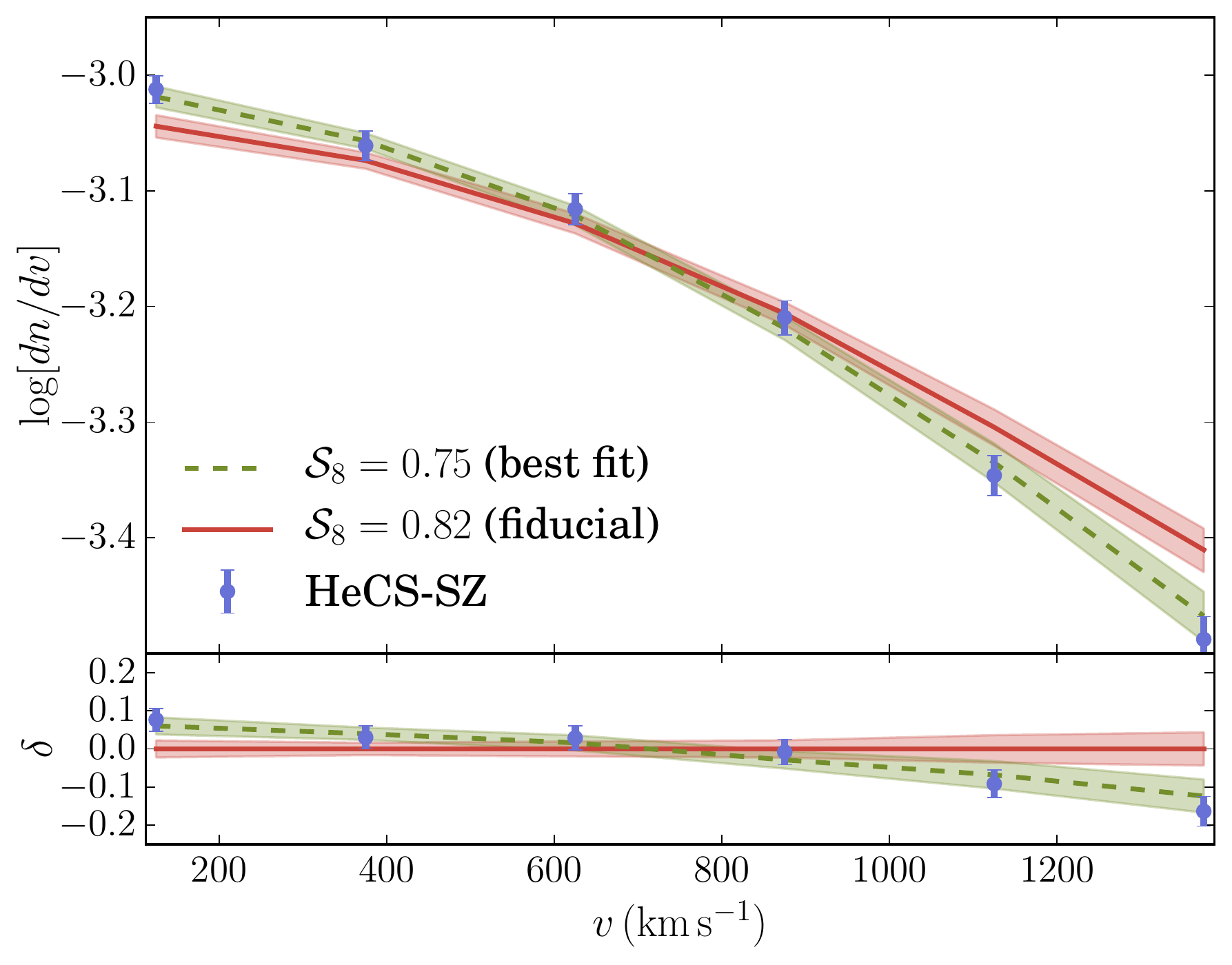}\\[5ex]
	Radial Selection & Rich Clusters & Reduced Interlopers \\
	\includegraphics[width=0.3\textwidth]{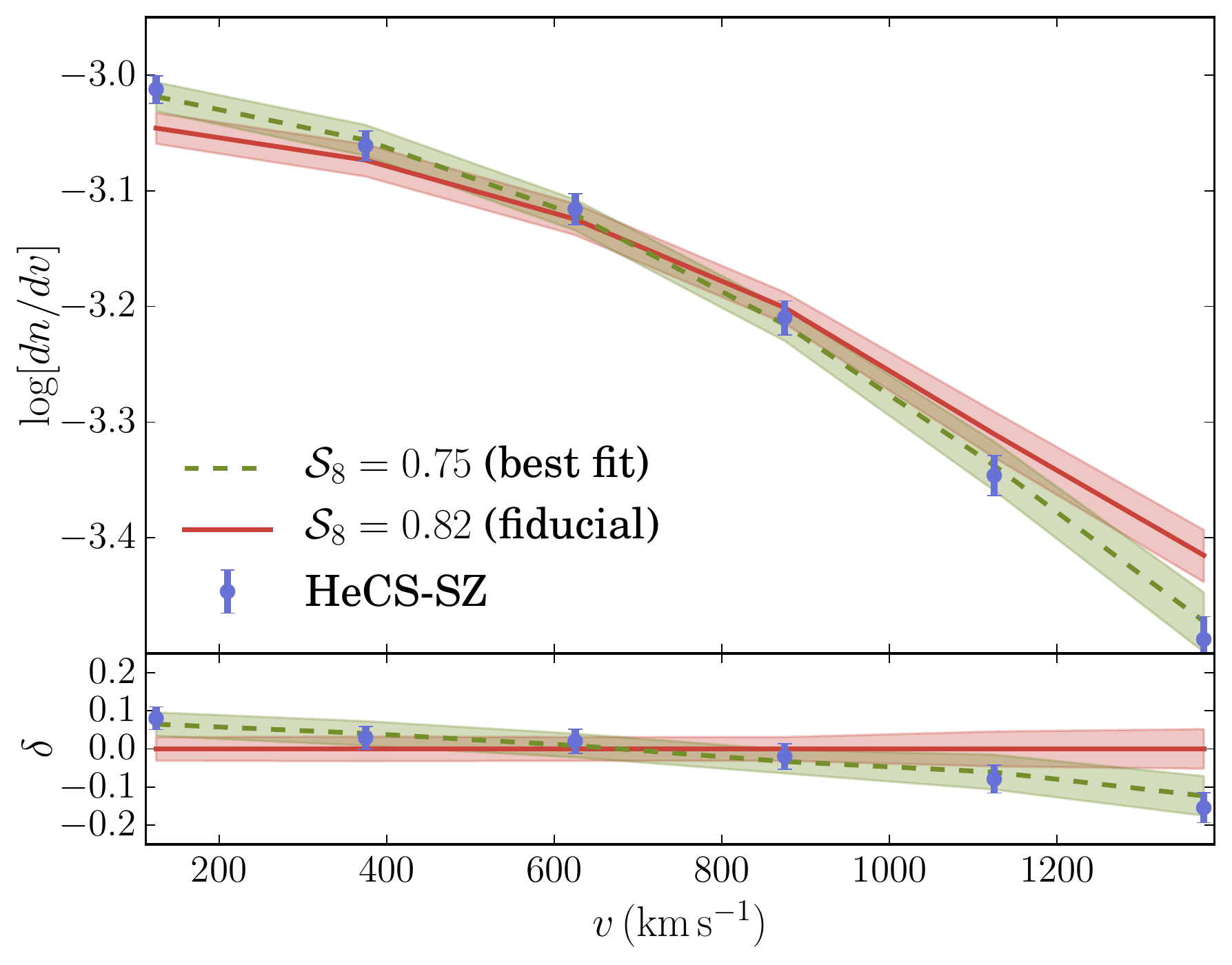}&\includegraphics[width=0.3\textwidth]{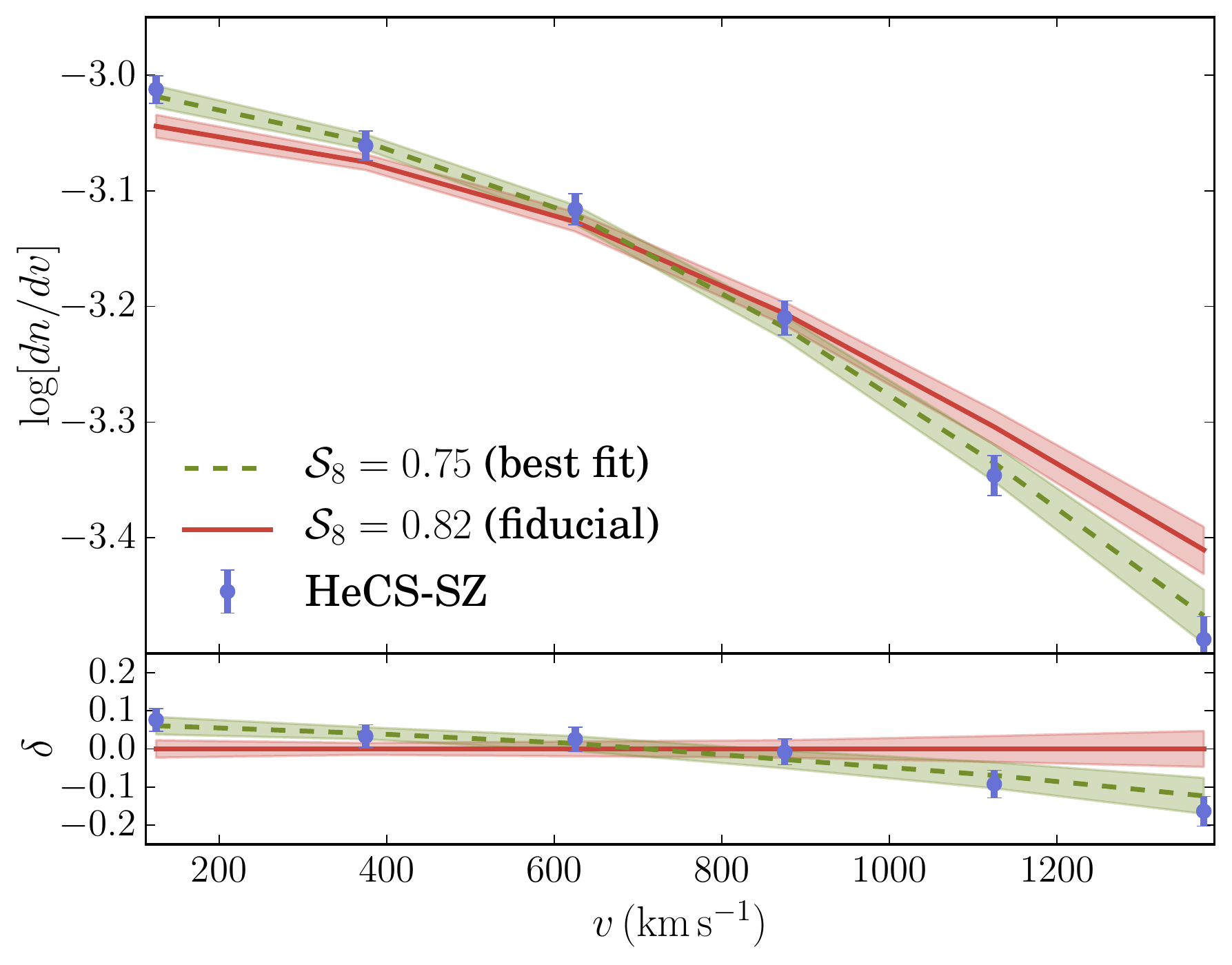}&\includegraphics[width=0.3\textwidth]{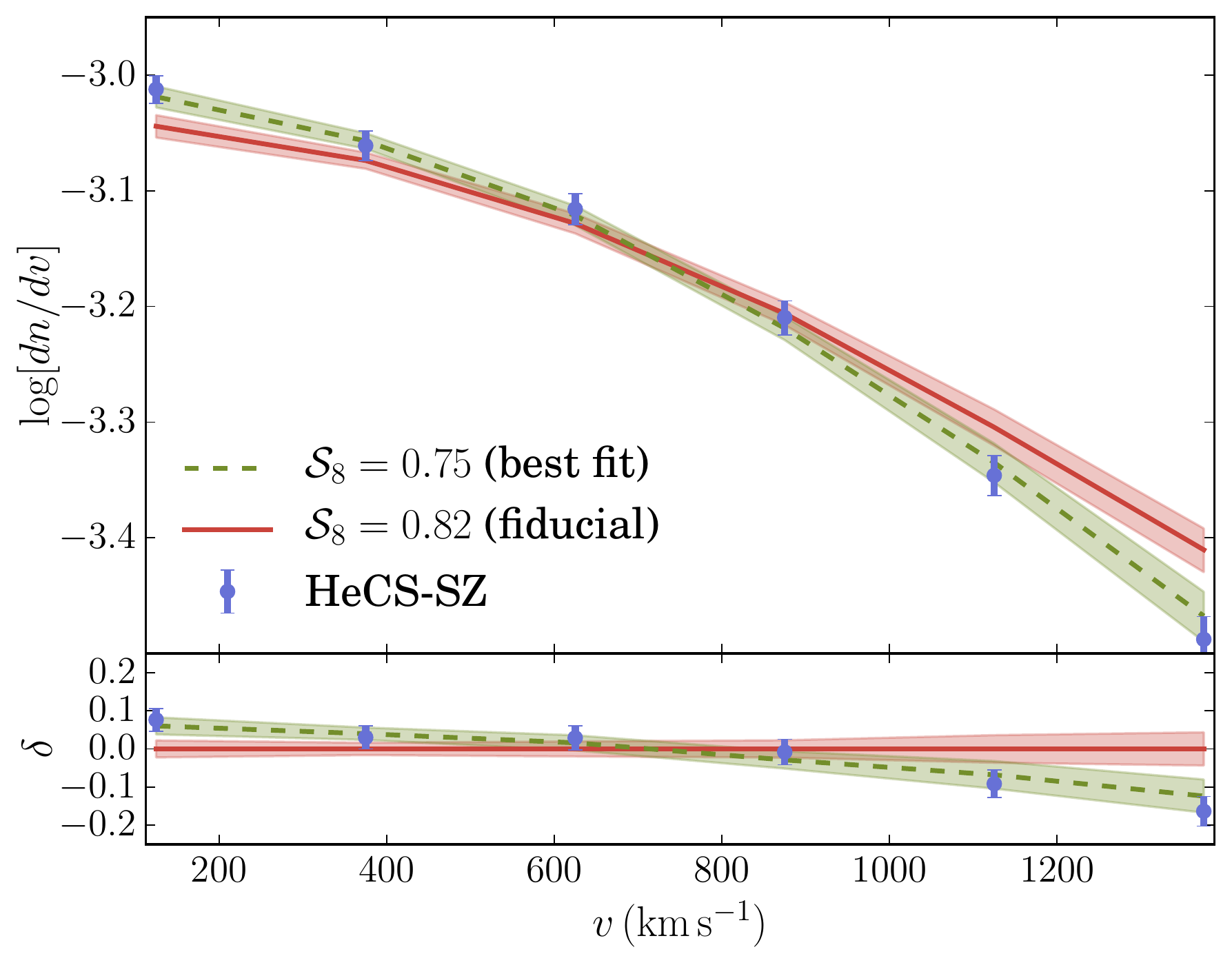}\\[5ex]
	Red Fraction & Quiescent Galaxies & Biased Velocities \\
	\includegraphics[width=0.3\textwidth]{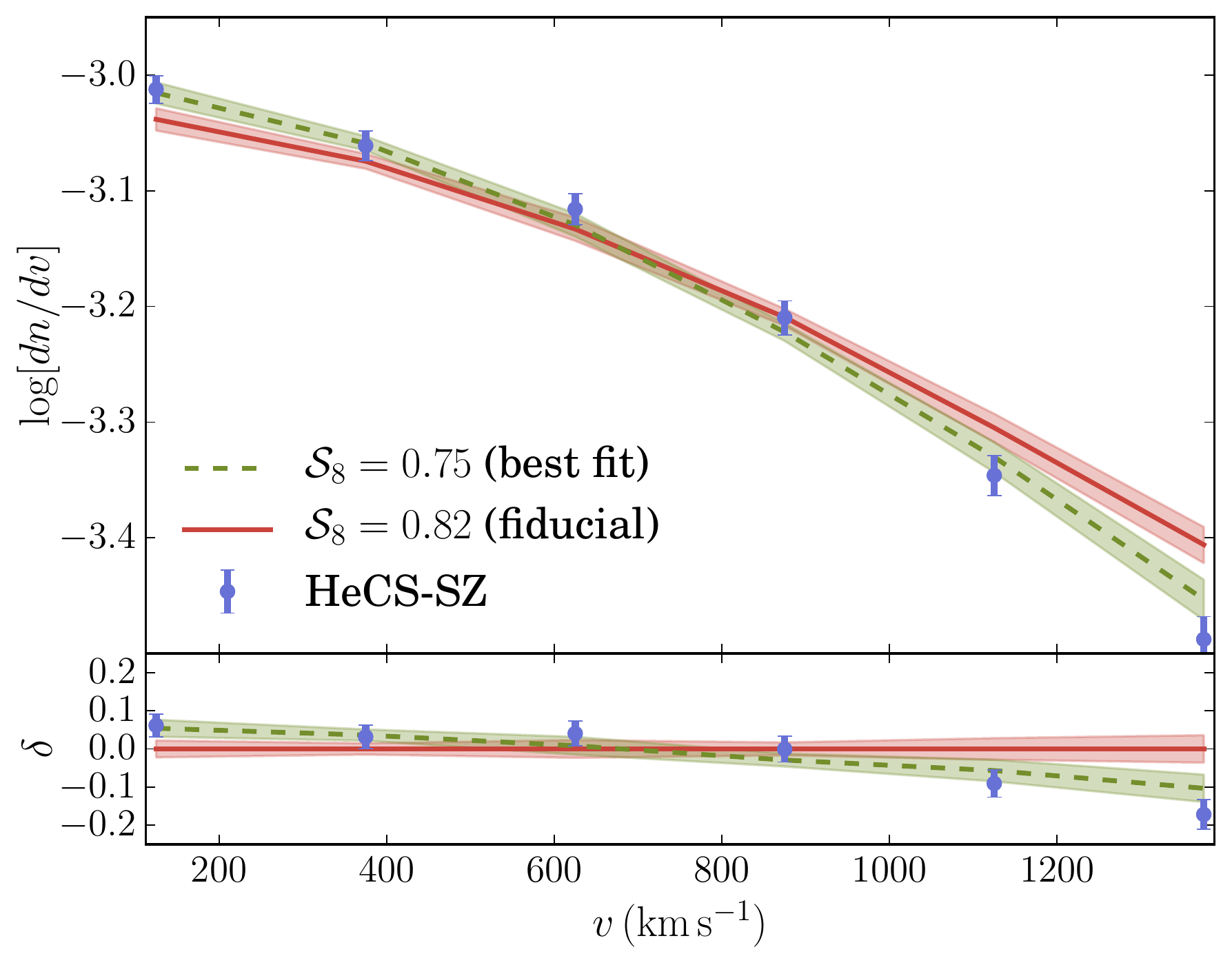}& \includegraphics[width=0.3\textwidth]{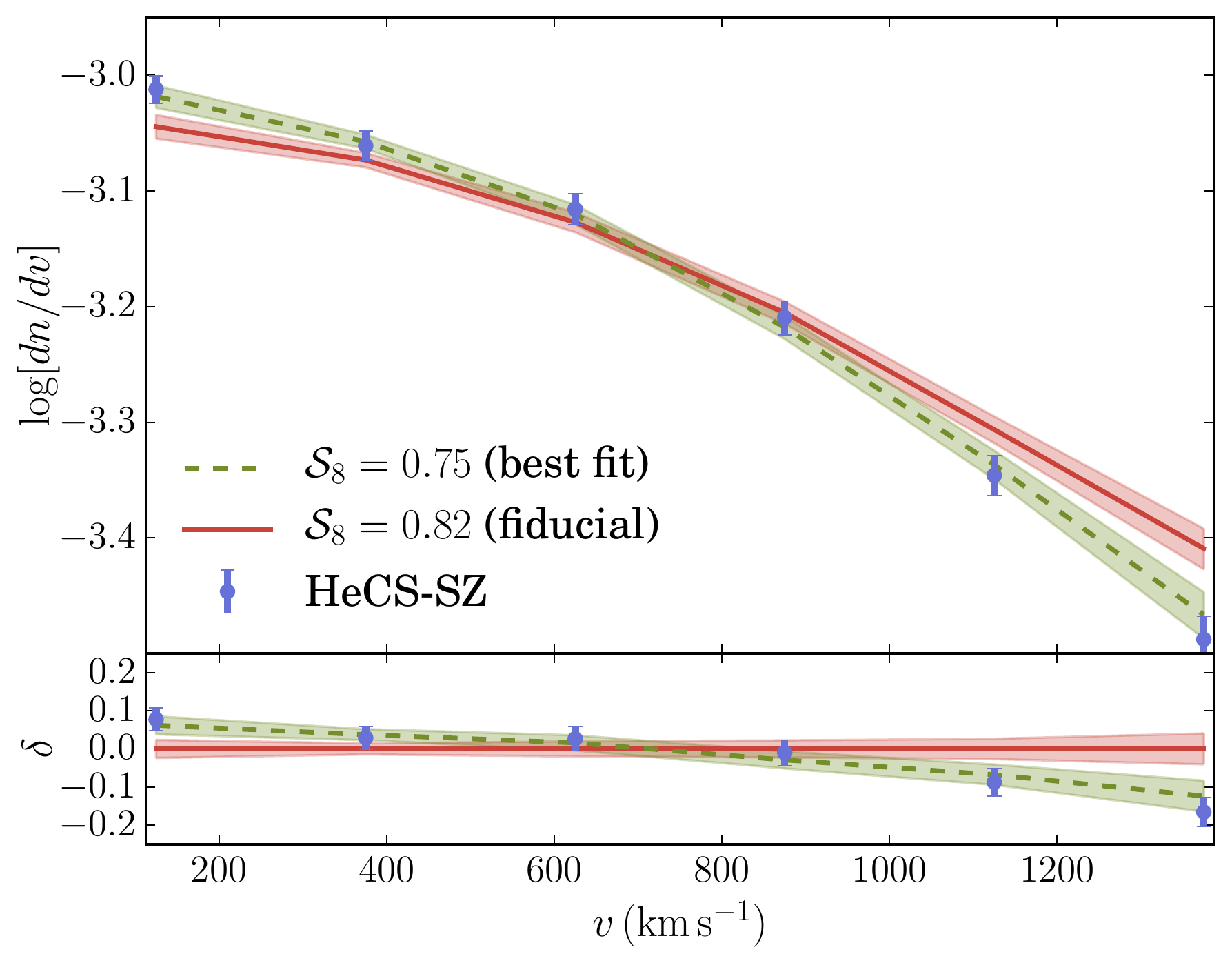} & \includegraphics[width=0.3\textwidth]{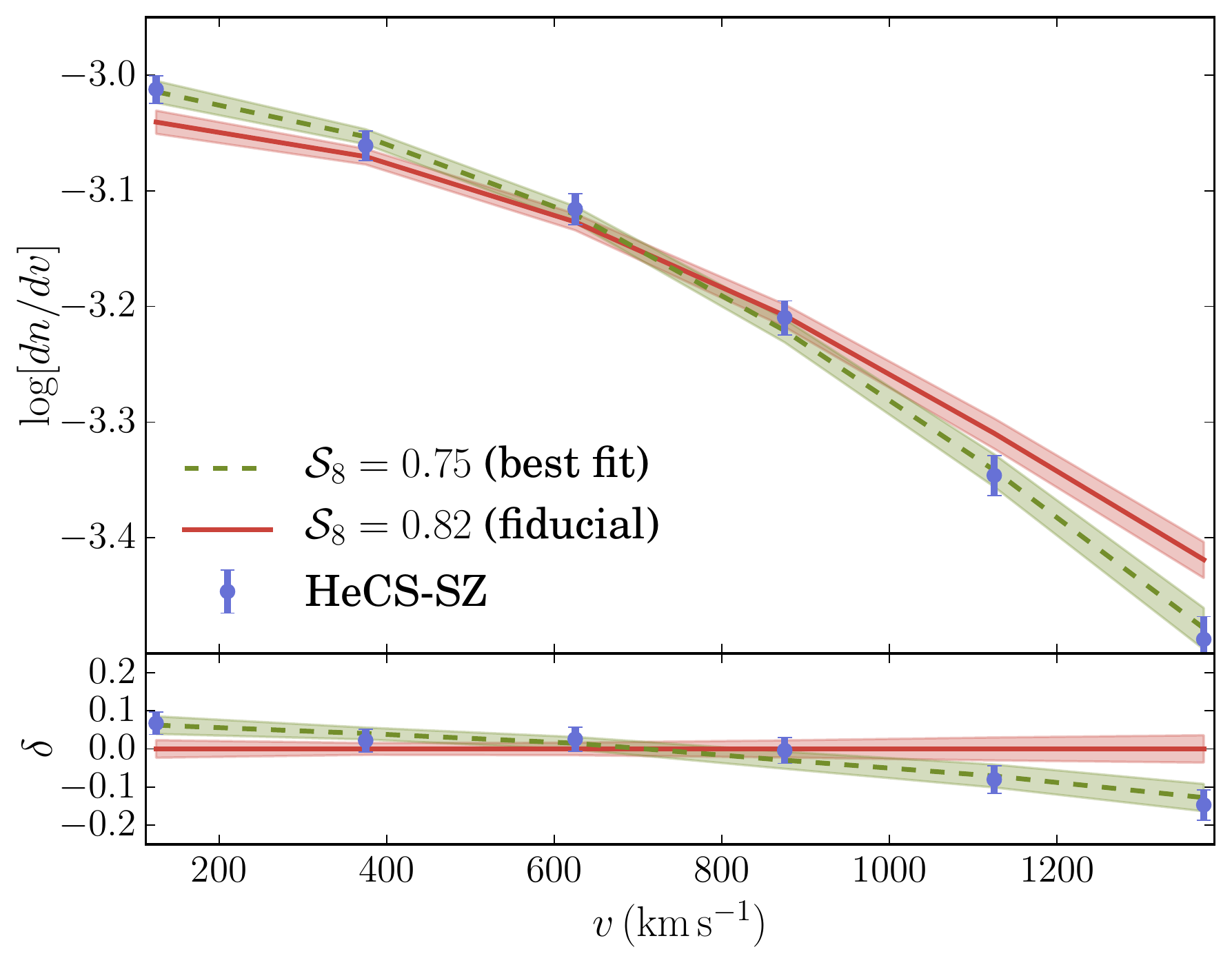}\\[5ex]
	
	\end{tabular}
	\end{center}
	\caption{{The VDF, plotted in the same format as in Figure \ref{fig:vdf}, for the mock catalogs with systematics.   The catalogs are described in detail in Section \ref{sec:altcat} and serve to explore how parameter choices and possible systematic differences between the \hecs sample and mock catalogs may affect the resulting constraints.  These have remarkably small effects:  the \hecs catalog (blue points with error bars) has fewer high-velocity galaxies compared to the fiducial mock catalog (red) and is in much better agreement with a lower-\Sig cosmology (shown is the standard catalog best fit catalog, green) for all of the alternative catalogs.}
	}
	\label{fig:vdf_alternate}
\end{figure*}

{Figure \ref{fig:vdf_alternate}  shows the VDF for mock catalogs with systematics, which  explore how parameter choices or possible biases between the \hecs sample and mock observations may introduce bias to the results.  For complete details on the parameter choices for each of the mock catalogs with systematics, see Section \ref{sec:altcat}.}

{Four of the mock catalogs with systematics (Large Aperture, No Axial Hole, Sigma Clipping, and Low Redshift) assess parameter choices.  For these catalogs, we change one of the standard parameter choices, and apply this change to both the mock catalog as well as the \hecs sample.  As can be seen in Figure \ref{fig:vdf_alternate}, both the mock catalog VDF as well as the \hecs VDF change compared to the standard version in Figure \ref{fig:vdf}.  This change is most pronounced in the sigma clipping example, with a decrease in signal at the high velocity end of the VDF.  In all cases, the \hecs VDF signal lies below (above) the fiducial mock VDF signal at high (low) velocities, and is in much closer agreement with a low \Sig cosmology.  }

The remaining {seven} mock catalogs with systematics {(Y-M Scatter, Radial Selection, Rich Clusters, Reduced Interlopers, Red Fraction, Quiescent Galaxies, and Biased Velocities)} involve changes to the mock VDF only; the \hecs VDF signal is the same as is shown for the standard catalog in Figure \ref{fig:vdf}.  The changes impart small changes on the mock VDF, but none of these changes are sufficient to pull the \hecs sample into agreement with the $\Sig=0.82$ fiducial cosmology VDF. {(There is one exception, the Biased Velocities catalog, which is within 1$\sigma$ of $\Sig=0.82$, but as discussed previously, this catalog likely overestimates a realistic velocity bias, with the Quiescent Galaxies catalog being a more proper modeling of this effect.)} The result is robust to these explorations of possible systematics:  the \hecs clusters are dynamically colder than those in the simulation at the fiducial cosmology.

\clearpage

\subsection{Radial Distribution Function}
\label{sec:rdf}

The radial distribution function (RDF) is used as a self-consistency check to evaluate the relative radial distributions of galaxies.  It is calculated as a sum of comoving radial distance PDFs and is given by
\begin{equation}
	\frac{dn}{dR_\mathrm{sep}} (R_\mathrm{sep}) = \left[\frac{1}{N}\sum_{i=1}^{N} \left[ \mathrm{PDF}(R_\mathrm{sep}) \right]_i \right]_{A, z_\mathrm{min}, z_\mathrm{max}},
	\label{eq:RDF}
\end{equation}
where $R_\mathrm{sep}$ is the {projected} radial distance from the galaxy to the cluster center, $\mathrm{PDF}(R_\mathrm{sep})$ denotes a probability distribution function of the projected comoving distance from the cluster center under the assumption of the fiducial cosmology, the index $i$ denotes a sum over $N$ clusters, and $A, z_\mathrm{min}, z_\mathrm{max}$ indicates that the RDF is calculated for a given sky area and redshift.

\begin{figure}[]
\begin{center}

	\includegraphics[width=0.5\textwidth]{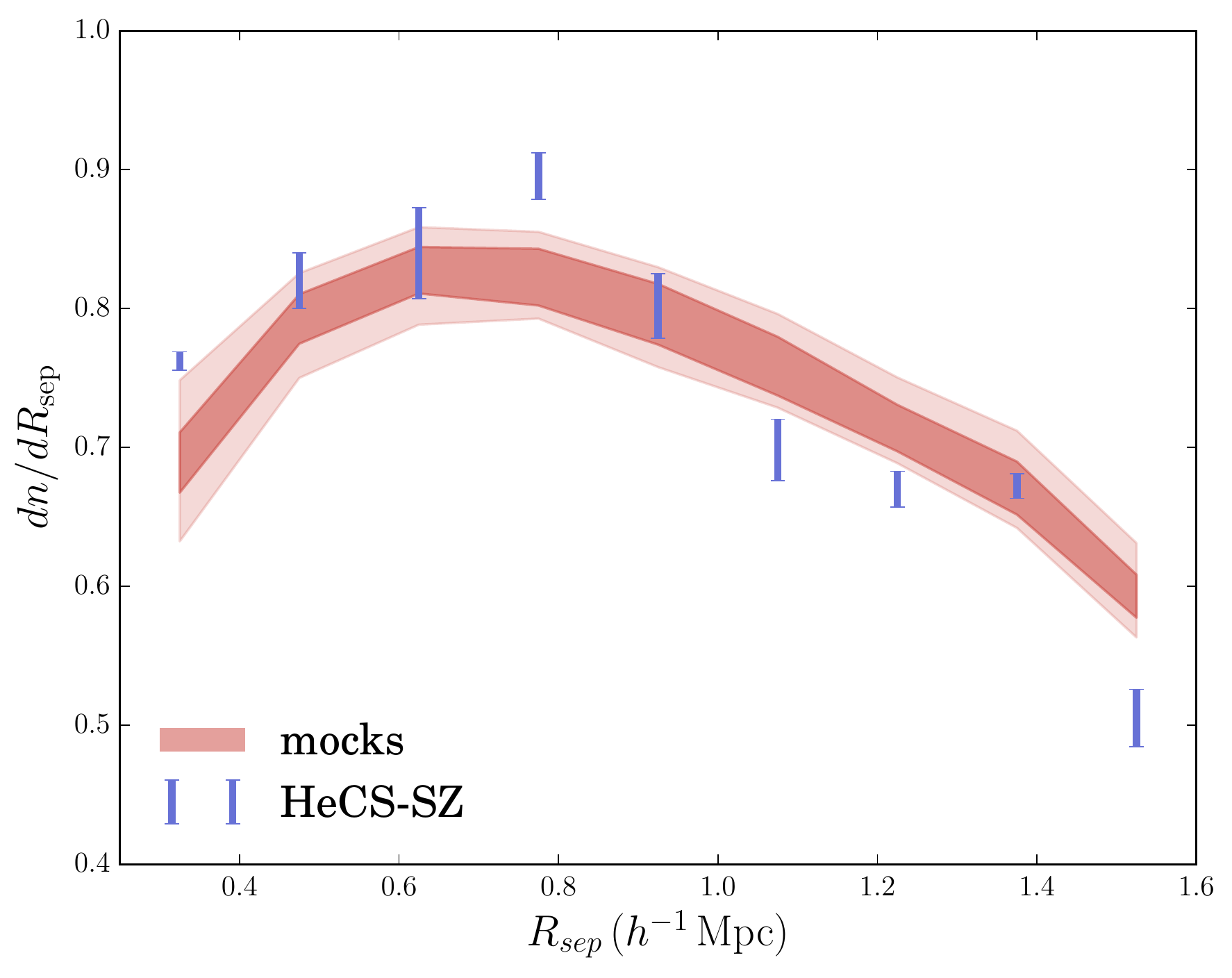}

\end{center}
	\caption{The radial distribution function (RDF) of the mock observations and \hecs clusters.  Parameters \om and $h$ affect the inferred $R_\mathrm{sep}$, and allowing these parameters to vary within the 2-$\sigma$ values reported by \cite{2016A&A...594A..13P} produces a band of $RDF$ curves for the mock observations (dark and light red bands showing the middle 68\% and 95\%, respectively).  The RDF of the \hecs sample (blue with Poisson error bars) is subtly more centrally peaked than the RDF of the mock observation.  {Subsampling the mock observation to match the \hecs RDF to account for possible galaxy selection effects is explored in the Radial Selection catalog (see Sections \ref{sec:altcat} and Table \ref{table:summary})}.  Alternately, the difference may be due to the \hecs sample containing less massive clusters with a smaller physical extent than the mock observation clusters.}

	\label{fig:rdf}
\end{figure}

Figure \ref{fig:rdf} shows the RDF for the mock clusters, which are based on the fiducial cosmology, and the \hecs clusters.  {The error bars on the \hecs RDF is created by allowing \om and $h$ to vary within an ellipse defined by the TT+lowP 2-$\sigma$ constraints reported by \cite{2016A&A...594A..13P}:  \om varies from \param{$0.29$ to $0.34$} and $h$ varies from \param{$0.65$ to $0.69$}.}  The error band on the RDF of the mock observations shows the middle 68\% and 95\% of the individual mock observation RDFs at the fiducial cosmology.  

When evaluated at the fiducial cosmology, the \hecs RDF is somewhat more centrally peaked than the mock observations, though the difference is subtle.  One explanation for this is that the \hecs sample may contain fewer massive clusters.  Alternately, galaxy selection effects, cluster selection effects, or baryonic effects may be the source of the  disagreement, which is not fully explained by allowing \om and $h$ to vary to extreme values.  The effect of galaxy selection is explored by subsampling the mock catalog to match the  \hecs sample RDF in the Radial Selection catalog described in Section \ref{sec:results} and Table \ref{table:summary}.

\section{{Results}}
\label{sec:results}

\subsection{Credible Regions}

The posterior probability, $P(\mathmodel{}|y)$, of a model given the \hecs observation is calculated as in  \cite{2017ApJ...835..106N}.  Briefly, the estimated covariance matrix, \smash{$\hat{C}$}, is given by
\begin{equation}
\hat{C}= \frac{1}{n_\mathrm{mock}-1} \sum_{i=1}^{n_\mathrm{mock}} \left[ (y_i-\bar{y})(y_i-\bar{y})^T\right],
\label{eq:cov}
\end{equation}
where $i$ denotes a sum over the $n_\mathrm{mock}=80$ fiducial mock observations, $n_\mathrm{bin}$ is the number of velocity bins, $y_i$ is a $n_\mathrm{bin}\times1$ column vector of the $i^{th}$ mock observation's $n_\mathrm{bin}$ bin values, and $\bar{y}$ is a $n_\mathrm{bin}\times1$ column vector of the \hecs observation. 

As detailed in, e.g., \cite{2007A&A...464..399H, 2013MNRAS.432.1928T, 2014MNRAS.439.2531P}, an unbiased estimator, $\hat{\Psi}^{-1}$, of the inverse covariance matrix is given by
\begin{equation}
\hat{\Psi}^{-1} = \frac{n_\mathrm{mock}-n_\mathrm{bin}-2}{n_\mathrm{mock}-1} \hat{C}^{-1},
\label{eq:invcov}
\end{equation}
where $\hat{C}^{-1}$ denotes a standard matrix inversion of the covariance matrix. The $\chi^2$ values are calculated by
\begin{equation}
\chi^2(y|\sigma_8, \Omega_m) = (\bar{y}-y^\star)^T \, \hat{\Psi}^{-1} \, (\bar{y}-y^\star),
\label{eq:chi2}
\end{equation}
{where $y^\star$ is a $n_\mathrm{bin}\times1$ column vector of average mock observation VDF signal at a single cosmology and $\bar{y}$ is the measured \hecs VDF signal.  The unbiased estimator of the inverse covariance matrix, $\hat{\Psi}^{-1}$, is calculated at the fiducial cosmology and assumed to be constant across the \sig-\om plane.}  This assumption breaks down as models farther from the fiducial model are considered.  To properly analyze the VDF and the constraints in the \model{} plane, one would need a suite of simulations across multiple \sig and \om models.

\begin{figure}[]
	\begin{center}
	\includegraphics[width=0.4\textwidth]{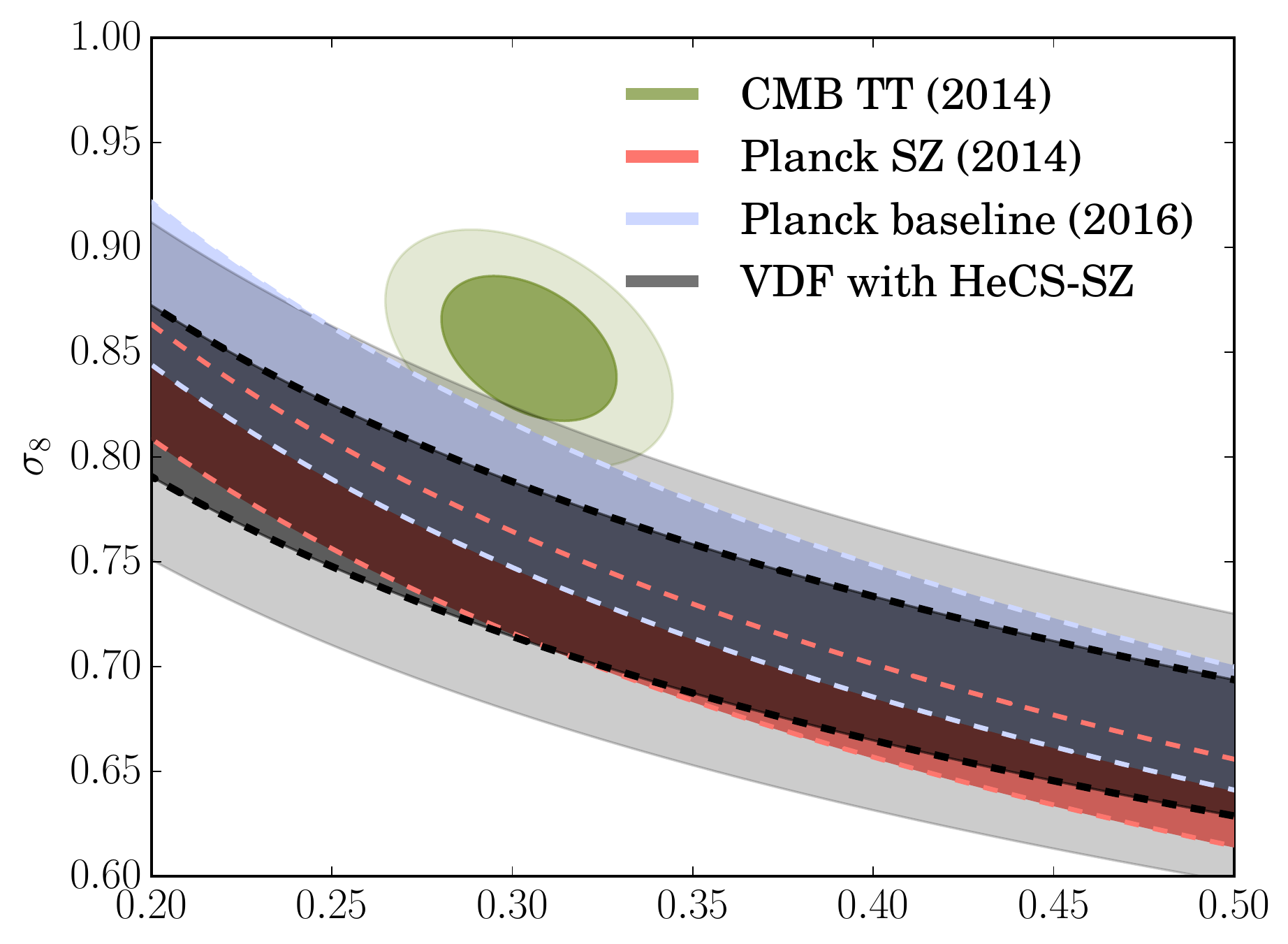} \\
	\includegraphics[width=0.4\textwidth]{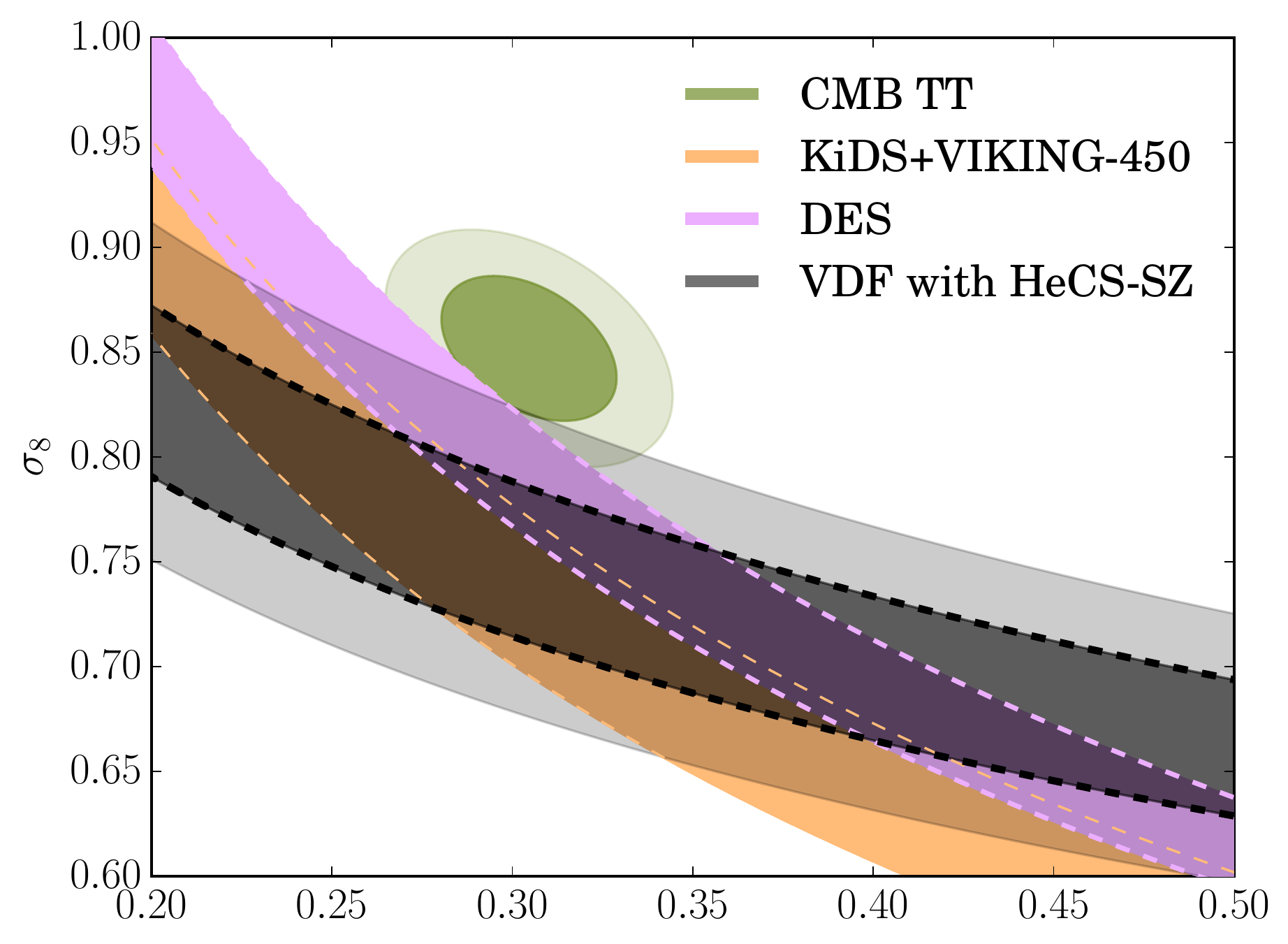} \\
	 \includegraphics[width=0.4\textwidth]{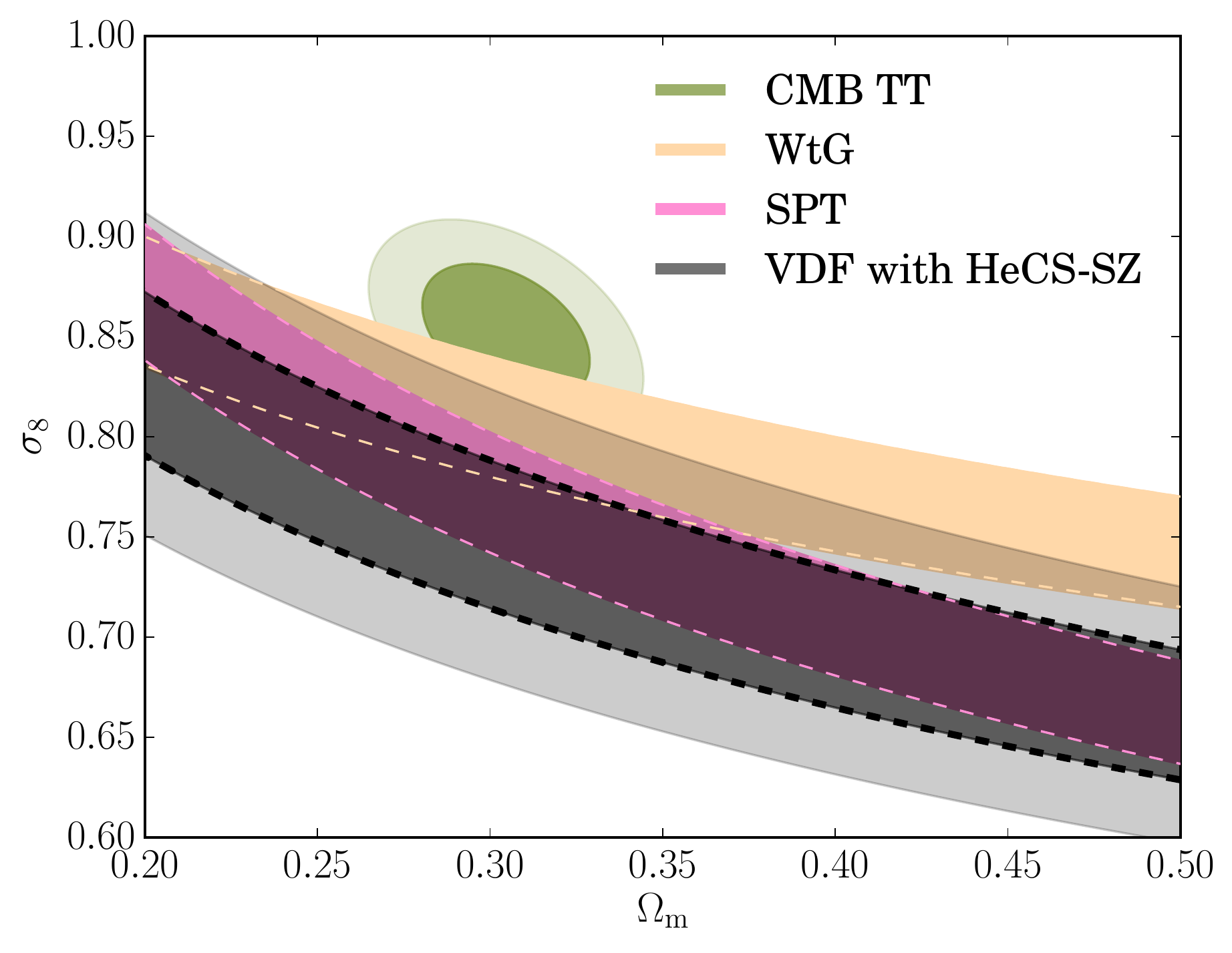}\\
	
	\caption{VDF cosmological constraints on \sig and \om (black, 68\% and 95\% contours, all plots).  
	The \planck CMB TT fiducial cosmology (green, 1- and 2-$\sigma$ contours, all plots) prefers a high \sig.
	\cite{2014A&A...571A..16P} constraints from \planck SZ cluster counts (red, top plot) prefers a low \sig if  mass bias is not invoked, while more recent {cluster-based} results invoking mass bias
	from \cite{2016A&A...594A..24P} CCCP + BAO + BBN [Baseline] (blue, top plot) lie below, though not in tension with, the \planck CMB fiducial cosmology.
	{The KiDS+VIKING-450 cosmic shear analysis \citep[orange, center plot,][]{2018arXiv181206076H}} also lies below the \planck CMB fiducial cosmology.  
	\cite{2017arXiv170801530D} Year 1 Results for $\Lambda CDM$ (purple, middle plot), 
	Weighing the Giants \citep[peach, bottom plot,][]{2015MNRAS.446.2205M}, 
	and
	South Pole Telescope \citep[pink, bottom plot,][]{2016ApJ...832...95D} 
	lie below, though not in tension with, the \planck fiducial cosmology.
	}
	\label{fig:likelihood}
	\end{center}
\end{figure}

We find that, for a given integral completeness $\mathcal{C}$, the likelihood $\mathcal{L}(\Sig|\mathcal{C})$ is well modeled by a Gaussian in \Sig with constant variance $\sigma^2$ and mean $\mu(\mathcal{C})$ that varies linearly with integral completeness,  $\mathcal{C}\equiv N_\mathrm{observed}/N_\mathrm{true}$.  

{To account for uncertainties in integral completeness, which was not an unknown quantity in \cite{2017ApJ...835..106N}, we add an additional step to the statistical inference by including priors on \Sig and $\mathcal{C}$.} The posterior $\pi$ takes the form
\begin{equation}
\pi(\Sig \mid \bar{y}, \mathcal{C}, \sigma) \propto \mathcal{L}\left( \Sig \mid \mathcal{C}, \sigma\right) \pi_1(\Sig) \pi_2(\mathcal{C})
\end{equation}
where the likelihood  is a Gaussian 
\begin{equation}
\mathcal{L}\left( \Sig \mid \mathcal{C}, \sigma\right) \propto \exp{\left(\frac{-\left[\Sig - \mu(\mathcal{C})\right]^2}{(2 \sigma)^2}\right)},
\end{equation}
$\pi_1(S_8)$ is a flat prior on \Sig, \Sig$\in[0.0, 1.0]$, and $\pi_2(\mathcal{C})$ is a flat prior on completeness,  $\mathcal{C}\in[0.6, 1.0]$.

 {The {credible regions} are calculated between $\om= 0.28$ and $\om=0.33$;} this function takes the form
\begin{equation}
	\mathcal{S}_8 = \sigma_8 \left( \frac{\Omega_m}{0.3}\right)^\gamma
\end{equation} 
where the power law parameter $\gamma$ describes the direction of the degeneracy in the \model{} plane and the parameter \Sig can be interpreted as the preferred value of \sig at $\Omega_m=0.3$.

Figure \ref{fig:likelihood} shows the 68\% and 95\% credible regions.  We find constraints in the \model{} plane given by \smash{$\sigma_8 \left(\Omega_m/0.3\right)^{0.25} = 0.751\pm0.037$}, as shown in Figure \ref{fig:likelihood}.  
The \planck CMB TT fiducial cosmology lies outside of our 95\% credible region, but 
constraints from \planck SZ cluster counts \citep{2014A&A...571A..16P},
{the KiDS+VIKING-450 cosmic shear analysis \citep{2018arXiv181206076H}} 
and DES Year 1 Results for $\Lambda CDM$ \citep{2017arXiv170801530D}  
lie within 1 $\sigma$ of our reported constraints.
More recent \planck cluster count results invoking mass bias \citep[baseline results,][]{2016A&A...594A..24P},
Weighing the Giants \citep{2015MNRAS.446.2205M},
and South Pole Telescope \citep{2016ApJ...832...95D}
have maximum likelihood functions that are within 2$\sigma$ of our reported constraints near $\om=0.3$.

Note that while some other definitions of \Sig use a $\gamma$ parameter of either $\gamma=0.5$ \citep[e.g.][]{2017MNRAS.465.1454H} or $\gamma=0.3$ \citep[e.g.][]{2014A&A...571A..20P}.  We allow $\gamma$ to be informed by the data rather than adopting a more standard definition, but for values of \om near the \om normalization value of $0.3$, these disparate definitions of \Sig can be compared.  It is only far from the \om normalization value of $0.3$ that the degeneracy direction $\gamma$ separates these different definitions of \Sig.

\begin{figure*}[]
	\begin{center}
	\includegraphics[width=0.8\textwidth]{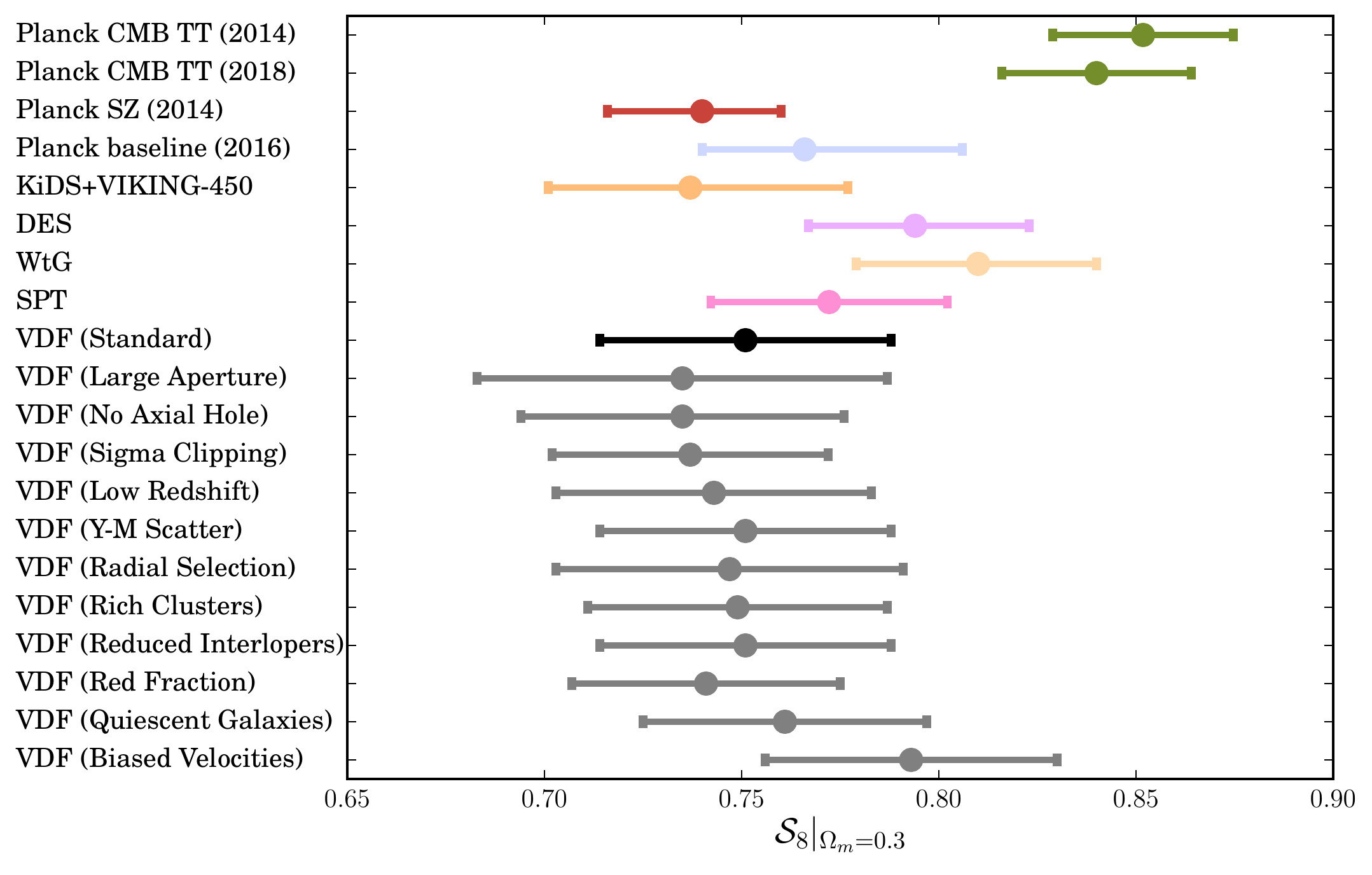} 
	\caption{{\Sig constraints evaluated at $\Omega_m=0.3$ for a variety of cosmological probes.  \planck CMB TT (2014), SZ, and baseline, as well as KiDS+VIKING-450 cosmic shear, DES, WtG, SPT, and VDF (Standard) are the same as are shown in Figure \ref{fig:likelihood}.  More recent \planck CMB TT results \citep{2018arXiv180706209P} are also shown in green.  {The Biased Velocities catalog (discussed in detail in Sections \ref{sec:altcat} and \ref{sec:altconstraints}) artificially imposes a fractional velocity bias on all galaxies in the mock catalog, likely overestimating the true velocity bias, and may overestimate \Sig constraints as a result. Constraints from the remaining catalogs with systematics show remarkable agreement, with central values within $\frac{1}{2}\sigma$ of the standard catalog. }}}
	\label{fig:s8summary}
	\end{center}
\end{figure*}

{
\subsection{\Sig Constraints for Mock Catalogs with Systematics}

\begin{deluxetable*}{l l r }

\tablecaption{{Catalog Summary}\label{table:summary}}
\tablehead{
\colhead{Approach} 	& \colhead{Description} & \colhead{Parameter Constraints}  	}

\startdata
Standard							& $R_\mathrm{ap}=1.6\Mpch$, \vcut$= 2500\,\kms $, $R_\mathrm{hole}=0.25\Mpch$, 			&\result{$\Sig = \sigma_8 \left(\Omega_m/0.3\right)^{0.25} = 0.751\pm0.037$}\\[1ex] 
								& $N_\mathrm{min}=20$															&\\[1.5ex]	
Large Aperture						& $R_\mathrm{ap}=2.3\Mpch$, \vcut $= 3785\, \kms$									&\result{$\Sig = \sigma_8 \left(\Omega_m/0.3\right)^{0.25} = 0.735\pm0.052$}\\[1.5ex] 
No Axial Hole						& $R_\mathrm{hole}=0.0\Mpch$													&\result{$\Sig = \sigma_8 \left(\Omega_m/0.3\right)^{0.25} = 0.735\pm0.041$}\\[1.5ex] 
Sigma Clipping						& $\sigma_\mathrm{clip}=2.0$														&\result{$\Sig = \sigma_8 \left(\Omega_m/0.3\right)^{0.25} = 0.731\pm0.035$}\\[1.5ex] 
Low Redshift 						&Limit catalogs to clusters with $z\leq0.2$.											&\result{$\Sig = \sigma_8 \left(\Omega_m/0.3\right)^{0.25} = 0.743\pm0.040$}\\[1.5ex] 
Y-M Scatter 						& 50\% additional scatter in Y-M relation.												&\result{$\Sig = \sigma_8 \left(\Omega_m/0.3\right)^{0.25} = 0.751 \pm0.037$}\\[1.5ex] 
Radial Selection 					& Force simulation RDF to match observed RDF by downsampling 							&\result{$\Sig = \sigma_8 \left(\Omega_m/0.3\right)^{0.25} = 0.747 \pm0.044$}\\[1ex] 
								& mock observation. 																&\\[1.5ex]	
Rich Clusters						& $N_\mathrm{min}=40$															&\result{$\Sig = \sigma_8 \left(\Omega_m/0.3\right)^{0.25} = 0.749\pm0.038$}    \\[1.5ex] 
Reduced Interlopers 					& Higher threshold for interlopers; interlopers must live in a halo  							&\result{$\Sig = \sigma_8 \left(\Omega_m/0.3\right)^{0.25} = 0.751\pm0.037$}    \\[1ex] 
								& with $M_\mathrm{200}\geq10^{12}\Msolarh$.											&\\[1.5ex]	
{Red Fraction} 					& Randomly select 80\% of true members and 20\% of interlopers 							&\result{$\Sig = \sigma_8 \left(\Omega_m/0.3\right)^{0.25} = 0.741\pm0.034$}      \\[1ex] 
								& to form the mock observation.													&\\[1.5ex]		
{Quiescent Galaxies} 			&Select only galaxies with sSFR $\leq10^{-1}\,\mathrm{Gyr}^{-1}$.							&\result{$\Sig = \sigma_8 \left(\Omega_m/0.3\right)^{0.25} = 0.761\pm0.036$}   \\[1.5ex] 
{Biased Velocities} 				&\new{Artificially bias the velocities of all mock galaxies.}								&\result{$\Sig = \sigma_8 \left(\Omega_m/0.3\right)^{0.25} = 0.793\pm0.037$}   \\[1.5ex] 
\enddata
\label{table:summary}
\end{deluxetable*}

\label{sec:altconstraints}

Figure \ref{fig:s8summary} shows a summary of \Sig constraints evaluated at $\om=0.3$.  This figure shows the sample of recent probes of \Sig that were highlighted in Figure \ref{fig:likelihood}, constraints of \Sig from the VDF standard catalog, and also constraints of \Sig for the mock catalogs with systematics.  Our constraints on \Sig are remarkably robust to the changes in catalog parameters that are explored by the \new{eleven} additional mock catalogs.}

{The Biased Velocities and Quiescent Galaxies catalogs constraints prefer the largest \Sig.  The Biased Velocities catalog (\result{$0.793\pm0.037$}, with a center that is $\approx 1.15 \sigma$ from the standard catalog), as discussed in Section \ref{sec:altcat}, overemphasizes the true velocity bias in two ways:  first, by imposing the bias on all galaxies in the sample, including interlopers, and second, by failing to disentangle the fact that this bias tends toward zero for well-sampled clusters \citep{2013ApJ...772...47S}.  Supporting the relevance of the second of these caveats, four clusters from the \hecs sample are studied in detail in \cite{2013ApJ...767...15R}, with the authors finding no significant bias in the estimates of velocity dispersions of red-sequence galaxies compared to the full galaxy sample. Though the value of $0.95$ may be an overestimate of the bias found in the \hecs cluster sample, it nonetheless provides an important cross-check for understanding how velocity bias might affect the resulting cosmological constraints.  }

{In the case of the Quiescent Velocities catalog (\result{$\Sig = 0.761\pm0.036$}), in selecting only the quiescent galaxies in the mock catalog, the number of high-velocity galaxies is somewhat reduced.  This results in constraints preferring a slightly larger \Sig, though still within $\frac{1}{2}\sigma$ of the standard method.  This can be understood in the context of velocity segregation \citep{2002A&A...387....8B, 2017MNRAS.468.1824O, 2018MNRAS.478.2618F}, with red or quiescent galaxies having a velocity dispersion that may be $\approx5\%$ smaller than that of the full cluster sample \citep{2005MNRAS.359.1415G,2006A&A...456...23B, 2013ApJ...773..116G, 2018MNRAS.481.1507B}, and can be attributed to the fact that blue galaxies are typically infalling and, therefore, have higher velocities.   However, it should be noted that the effect becomes smaller with well-sampled clusters \citep{2013ApJ...772...47S} and recent simulations find a stellar-mass-selected sample to be relatively unbiased ($<5\%$ in \cite{2018MNRAS.474.3746A}), and both of these may contribute to the fact that the Quiescent Velocities constraints are not as extreme as the Biased Velocities constraints.}

{Table \ref{table:summary} briefly describes the mock catalogs with systematics and also gives the parameter constraints on \Sig for each.
Regardless of the choice of parameters,  each approach yields a preference for a low \Sig and {all catalogs (except for the Biased Velocities catalog) have a central value within $\frac{1}{2}\sigma$ of the standard method.  While properly assessing the systematic error due to the choice of modeling parameters would require more thoroughly sampling the high-dimensional space of these parameters, this result suggests that the systematic error is of the order one-half the statistical error.}
}

\section{Discussion \& Conclusion}
\label{sec:discussion}

The abundance of clusters as a function of mass and redshift is a valuable tool for constraining cosmological models.  Notably, the cosmological parameters based on \planck CMB observations predict more high-mass clusters than are observed.  Biases in cluster mass estimates remain at the forefront of the discussion in interpreting cluster mass observations and resolving the tension between \planck CMB cosmological parameters and cluster counts. 

Eddington bias, caused by the steeply declining mass function coupled with errors in mass estimates, produces an observed halo mass function with an upscatter of high-mass clusters.  Accounting for this bias must be done correctly, and assumptions about the distribution of scatter handled carefully.  The velocity distribution function (VDF) is a forward-modeled test statistic that can be used to quantify the abundance of galaxy clusters in a way that is less sensitive to biases introduced by measurement error than more standard HMF approach.  

We have used the VDF to compare the summed velocity PDFs of observed \hecs clusters to mock observations produced by $N$-body simulations.  In agreement with  \cite{2016ApJ...819...63R},  we find that this collection of clusters are dynamically colder than expected, having  smaller velocities than one would predict for a \planck-selected sample of clusters for the \planck fiducial cosmology\new{.  This suggests} that the observed clusters may be less massive than an $\Sig=0.82$ cosmology would predict.  {We have explored several possible sources of systematic error, including parameter choices, interloper fraction, {velocity bias}, and galaxy selection effects.  Our results are remarkably robust to reasonable changes to the standard mock catalog.  }While the precise fit of our preliminary constraints has a small dependence on the details of the model, all approaches show a preference for a low \Sig, and our standard approach gives credible regions in the \model{} plane given by \result{$\Sig \equiv \sigma_8 \left(\Omega_m/0.3\right)^{0.25} = 0.751\pm0.037$}.  

The constraints presented here should not be overinterpreted.  The missing high-velocity members may be caused by a true dearth of high-mass clusters, or may alternately be caused by a bias between simulated cluster substructure velocity and cluster member velocities.  Also, assumptions of a smoothly-varying covariance matrix break down far from the fiducial model. {Similarly, the fraction and characteristics of interlopers are a function of cosmological parameters; these will also change far from the simulated fiducial cosmology.}  Properly evaluating the covariance matrices {and interloper population} of non-fiducial (or non-simulated) cosmologies would require a suite of large volume, high resolution simulations to  capture the nuanced correlations.  {Furthermore, properly evaluating these with realistic galaxy selection effects may require a large suite of hydrodynamical simulations that properly model galaxy properties.}  Such a suite of simulations could also eliminate the need to assume cosmological parameters in the calculation of comoving distances, which could instead be replaced with invariant parameters such as angular extent.

Other LSS and cluster-based analyses have found similar results, seeing a tension between observations and the \planck CMB cosmology, and preferring smaller \Sig values (or needing to invoke new physical models) to explain the discrepancy  \citep[e.g.][]{2013MNRAS.432.2433H, 2017MNRAS.465.1454H, 2017MNRAS.467.3024L, 2017arXiv170809813L}.  The VDF provides a complementary test to use clusters as a cosmological probe.  As observations probe larger areas of the sky to deeper magnitudes become available, these data sets will provide opportunities to understand and resolve the tension in \Sig constraints.

\acknowledgments{ 
{We thank our anonymous referee for their helpful feedback on this manuscript.} We also thank Peter Behroozi, Risa Wechsler, Andrew Hearin, and Charlie Conroy for early access to the UniverseMachine Catalog as well as {Sownak Bose,} Jessi \new{Cisewski-Kehe}, Daniel Eisenstein, Gus Evrard, Arthur Kosowsky,  Rachel Mandelbaum, Frank van den Bosch, \new{and Alexey Vikhlinin} 
for their valuable feedback on this project.
Hy Trac is supported in part by DOE grant DE-SC0011114 and NSF grant IIS-1563887. 
The CosmoSim database used in this paper is a service by the Leibniz-Institute for Astrophysics Potsdam (AIP).
The MultiDark database was developed in cooperation with the Spanish MultiDark Consolider Project CSD2009-00064.
The Bolshoi and MultiDark simulations have been performed within the Bolshoi project of the University of California High-Performance AstroComputing Center (UC-HiPACC) and were run at the NASA Ames Research Center. The MultiDark-Planck (MDPL) and the BigMD simulation suite have been performed in the Supermuc supercomputer at LRZ using time granted by PRACE.\\ }

\bibliography{../../references}
\bibliographystyle{apj}

\end{document}